\documentclass[prd,onecolumn,aps,nofootinbib,eqsecnum,notitlepage,nobibnotes,superscriptaddress,preprintnumbers]{revtex4-1}

\usepackage[utf8]{inputenc}
\usepackage{color}
\usepackage{amsfonts}
\usepackage{graphicx}
\usepackage[dvipsnames]{xcolor}
\definecolor{refs}{RGB}{245,156,74}
\usepackage[colorlinks=true,hyperfootnotes=true,citecolor=cyan]{hyperref}
\usepackage{dsfont} 
\usepackage{amsmath}

\usepackage{here}
\usepackage{graphicx}
\usepackage{amsmath,amsthm,amssymb}
\usepackage{bm}
\usepackage{color}

\usepackage{amsfonts}
\usepackage{dcolumn}
\usepackage{hyperref}
\allowdisplaybreaks[1]
\usepackage{ulem}

\begin{document}
\newcommand{\newc}{\newcommand}

\newcommand{\rk}{\textcolor{red}}
\newc{\ben}{\begin{eqnarray}}
\newc{\een}{\end{eqnarray}}
\newc{\be}{\begin{equation}}
\newc{\ee}{\end{equation}}
\newc{\ba}{\begin{eqnarray}}
\newc{\ea}{\end{eqnarray}}
\newc{\D}{\partial}
\newc{\rH}{{\rm h}}
\newc{\rd}{{\rm d}}
\newc{\rN}{{\rm N}}
\newc{\cX}{{\cal X}}
\newc{\C}{{\cal C}}
\newc{\X}{{\cal X}}
\newc{\Y}{{\cal Y}}
\newc{\rs}{{\rm s}}
\newc{\tN}{{\tiny {\rm N}}}

\title{Cosmological perturbations for ultra-light axion-like particles \\
in a state of Bose-Einstein condensate}

\author{Shinji Tsujikawa}
\email{tsujikawa@waseda.jp}
\affiliation{Department of Physics, Waseda University, 3-4-1 Okubo, 
Shinjuku, Tokyo 169-8555, Japan.}

\begin{abstract}

For ultra-light scalar particles like axions, dark matter can form a state of 
the Bose-Einstein condensate (BEC) with a coherent classical wave
whose wavelength is of order galactic scales. 
In the context of an oscillating scalar field with mass $m$, 
this BEC description amounts to integrating out the field oscillations 
over the Hubble time scale $H^{-1}$ in the regime $m \gg H$.
We provide a gauge-invariant general relativistic framework for studying 
cosmological perturbations in the presence of  
a self-interacting BEC associated with a complex scalar field.
In particular, we explicitly show the difference of BECs from 
perfect fluids by taking into account cold dark 
matter, baryons, and radiation as a Schutz-Sorkin 
description of perfect fluids.
We also scrutinize the accuracy of commonly used Newtonian 
treatment based on a quasi-static approximation for 
perturbations deep inside the Hubble radius. 
For a scalar field which starts to oscillate after matter-radiation equality, 
we show that, after the BEC formation, a negative self-coupling
hardly leads to a Laplacian instability of the BEC density contrast.
This is attributed to the fact that the Laplacian instability does not 
overwhelm the gravitational instability for self-interactions 
within the validity of the nonrelativistic BEC description.
Our analysis does not accommodate the regime of parametric resonance 
which can potentially occur for a large field alignment 
during the transient epoch prior to the BEC formation.

\end{abstract}

\date{\today}


\maketitle

\section{Introduction}
\label{introsec}

There are compelling observational evidences that 
about 25\,\% of today's energy density of the Universe 
is made of dark matter (DM). 
The existence of DM has been probed by temperature 
anisotropies in the Cosmic Microwave 
Background (CMB) \cite{WMAP,Planck}
as well as by the galaxy-clustering surveys ranging from 
large-scale superclusters down to small-scale dwarf 
galaxies \cite{Tegmark,Ho:2012vy,Betoule:2014frx}. 
To reveal the origin of DM is one of the most challenging 
problems in modern cosmology and particle physics. 

{}From the theoretical viewpoint, ultra-light bosons like axions 
can be good candidates for DM \cite{Baldeschi:1983mq,Sin:1992bg}. 
The axion is a psedo-Nambu Goldstone boson originally introduced 
to address the strong CP problem in quantum 
chromodynamics (QCD) \cite{Peccei:1977hh,Kim:1979if,Shifman:1979if}.
The mass of QCD axions is in the range
$m \gtrsim 10^{-6}$~eV to avoid overclosing the Universe \cite{Preskill:1982cy}.
String theory also gives rise to axions as Kaluza-Klein zero modes of 
anti-symmetric form fields \cite{Witten:1984dg,Svrcek:2006yi}. 
Depending on the geometry of string-theory 
compactification, the mass of axions can span over the light range 
from $10^{-33}$~eV to $10^{-10}$~eV \cite{Arvanitaki:2009fg}.
In such cases, the axion affects the late-time cosmological dynamics 
as all or a part of DM \cite{Amendola:2005ad,Hlozek:2014lca}. 
For the mass $m \sim 10^{-33}$~eV,  
the axion potential energy can even work as 
dark energy (DE) \cite{Kim:1998kx,Choi:1999xn,Nomura:2000yk,Kim:2002tq,Panda:2010uq}.

Cosmologically, the axion field $\phi$ is nearly frozen up to the instant where the 
Universe expansion rate $H$ drops below 
its mass $m$ \cite{Preskill:1982cy,Abbott:1982af,Dine:1982ah,Kim:1986ax,Kim:2014tfa,Marsh:2015xka}.  
The moment at which the axion starts to oscillate around the minimum of 
the potential $V(\phi)=m^2 \phi^2/2$ can be quantified by the condition 
$m \simeq 3H$ \cite{Hlozek:2014lca,Marsh:2015xka}. 
After many times of oscillations, the axion
field behaves as nonrelativistic matter 
in the form of a Bose-Einstein condensate (BEC) \cite{Sikivie:2009qn}.
In this BEC state, the bosonic particles behave as a classical coherent wave 
with the Compton wavelength $\sim m^{-1}$. 
For the ultra-light axion mass mentioned above, 
the Compton wavelength can reach the 
galactic scales and hence there is an intriguing possibility for 
probing observational signatures of 
such ``fuzzy DM'' \cite{Hu:2000ke,Hui:2016ltb}.

Indeed, the BEC has a ``quantum pressure'' which works against the 
gravitational clustering below a certain scale 
$\lambda_J$ \cite{Lif,Khlopov:1985jw,Hu:2000ke,Hwang:2009js,Marsh:2010wq}.
This Jeans scale corresponds to the de Broglie wavelength of 
a particle in the BEC ground state. 
For scales below $\lambda_J$ the quantum pressure manifests itself by 
the uncertainty principle, while the BEC density perturbation 
on scales larger than $\lambda_J$ grows as in the standard 
Cold-Dark-Matter (CDM). 
For the mass range around $m=10^{-22}$\,eV\,$\sim\,$10$^{-21}$~eV, 
the BEC DM can suppress the small-scale matter power 
below the 1~Kpc scale \cite{Hu:2000ke}. 
This allows a possibility for alleviating the excess of abundances
of dwarf galaxies present in the CDM model. 
The statistical analysis of Refs.~\cite{Irsic:2017yje,Armengaud:2017nkf} 
using Lyman-$\alpha$ forest data of the small-scale matter power spectrum 
placed the $2\sigma$ bound $m>2 \times 10^{-21}$~eV. 
On the other hand, there is a claim that the mass of order 
$10^{-22}$~eV is still allowed due to uncertainties 
in a thermal state of the high-redshift intergalactic 
medium \cite{Zhang:2017chj}.

For the smaller axion mass $m \lesssim 10^{-27}$~eV, 
the scalar field starts to oscillate after 
matter-radiation equality \cite{Hlozek:2014lca,Marsh:2015xka}.
In this case, the Jeans scale $\lambda_J$ can be within
the observable range of 
linear matter power spectrum and CMB temperature anisotropies. 
Since the quantum pressure suppresses the gravitational instability of
the BEC density contrast for scales smaller than $\lambda_J$, 
the axion field can not be all DM in this ultra-light mass region. 
Indeed, for the mass $10^{-32}~{\rm eV} \le m \le 10^{-25.5}~{\rm eV}$, 
today's axion density parameter is constrained to be less than 5\,\% 
of all DM \cite{Hlozek:2014lca}. 
Even with such a small density parameter, 
the axion coupling to photons gives rise to an interesting possibility 
for explaining the isotropic birefringence \cite{Fujita:2020aqt,Fujita:2020ecn} 
recently reported by analyzing the Planck2018 
data \cite{Minami:2020odp} (see also Refs.~\cite{Carroll:1998zi,Lue:1998mq}).

The standard BEC has a self-interaction whose effective potential is 
related to the s-scattering length. For axions, expanding the periodic 
potential $V(\phi)=m^2 f^2 [1-\cos(\phi/f)]$ around $\phi=0$ gives rise to 
an effective self-coupling energy density $\lambda_s \phi^4$ 
with $\lambda_s=-m^2/(24f^2)$.
Since the coupling constant $\lambda_s$ is negative for axions, 
this leads to an attractive self-interaction.
In a self-gravitating system of BEC, there is a possibility that 
this attractive force enhances the gravitational instability of BEC.
The effects of self-interactions on the dynamics of BEC perturbations 
and the formation of boson stars have been studied in 
Refs.~\cite{Chavanis:2011zi,Chavanis:2011uv,Chavanis:2011zm,Erken:2011dz,Guth:2014hsa,Eby:2016cnq,Suarez:2016eez,Levkov:2016rkk,Helfer:2016ljl,Zhang:2017flu,Cedeno:2017sou,Zhang:2017dpp,Desjacques:2017fmf,Suarez:2017mav,Arvanitaki:2019rax}.
For attractive interactions there are some particular scales in which 
the effective sound speed squared of linear BEC cosmological 
perturbations becomes negative, which can induce Laplacian 
instabilities.

Most of the works about the BEC cosmological perturbations 
in the literature have been based on the nonrelativistic Gross-Pitaevskii-Poisson (GPP) 
equation for a single wave function \cite{Nishiyama:2004ju,Boehmer:2007um,Fukuyama:2007sx,Harko:2011jy,Chavanis:2011uv}. 
The GPP equation, which follows from a Hamiltonian of interacting 
condensed bosons, corresponds to the Newtonian limit of 
the self-gravitating nonrelativistic BEC \cite{Dalfovo:1999zz}. 
With a Madelung representation of the wave function \cite{Madelung}, 
the BEC perturbation equations of motion can be expressed 
in terms of Newtonian analogue of the continuity and 
Euler equations. It is not yet clear whether this Newtonian 
approach remains valid for large-scale perturbations where 
the general relativistic effect on the dynamics of inhomogeneities
comes into play.

In this paper, we take the full general relativistic, covariant approach to 
the study of cosmological perturbations for nonrelativistic BEC. 
We begin with an explicit Lagrangian of a complex massive scalar field $\chi$ 
preserving a $U(1)$ charge, with its self-interaction involved. 
The similar treatment of BECs with the Lagrangian description was performed 
in Refs.~\cite{Fagnocchi:2010sn,Bettoni:2013zma,Ivanov:2019iec}, 
but the full relativistic treatment including the effect on metric 
perturbations was not fully addressed yet.
We also take into account CDM, baryons, and radiation as perfect fluids
to accommodate the case in which the BEC is not responsible 
for all DM. 
We clarify the difference between BEC and perfect fluids in terms of 
the covariant description of continuity and Euler equations.
We derive the full linear perturbation equations without fixing any 
particular gauge conditions and express them in terms of gauge-invariant variables. 
Thus, these master equations can be applied to any convenient gauge choices at hand.

We also study how the perturbation equation of the BEC 
density contrast can be recovered in the Newtonian limit. 
Even though the scalar-field oscillation 
is integrated out at the background level by the Madelung transformation, 
this oscillating mode appears in the perturbation equations.
However, we show that the existence of this mode hardly affects 
the dynamics of matter perturbations.
The approximate second-order equation for the density contrast, 
which we will derive in the Newtonian limit with the neglect of 
the oscillating mode, is in good agreement with the full numerical solution 
except for wavelengths close to the Hubble radius.
In addition, we will see that the density contrasts of perfect fluids like CDM 
and baryons are affected by the BEC sound speed $c_s^2$ through their 
gravitational interactions with BEC. 

We discuss the effect of BEC self-interactions on the dynamics 
of density perturbations as well. 
For the mass range $m=10^{-22}$\,eV\,$\sim\,$10$^{-21}$~eV 
within which the axion can be the source for all DM, 
the axion self-interaction is relevant to the growth of linear 
perturbations only in the deep radiation 
era \cite{Desjacques:2017fmf}. In this case, there are particular scales 
around 1~Mpc in which the self-interaction dominates over the 
quantum pressure.
For the ultra-light mass range $m \lesssim 10^{-27}$~eV, the self-interaction 
can be important on larger scales which are in the observational range of 
CMB and linear matter spectra. 
In this latter case, we will study its effect on the dynamics 
of BEC and CDM/baryon density contrasts during matter dominance. 
We show that, in spite of a negative value of $c_s^2$ for 
some particular scales, the Laplacian instability of BEC perturbations is 
suppressed relative to the gravitational instability in the regime where 
the nonrelativistic BEC description is valid. 
Since this description amounts to averaging over the field oscillations 
during the Hubble time scale, 
it does not accommodate the transient epoch toward the BEC formation 
during which parametric resonance can potentially enhance the BEC 
density contrast \cite{Zhang:2017flu,Cedeno:2017sou,Zhang:2017dpp,Arvanitaki:2019rax}.

Throughout the paper, we use the natural unit where the speed of light $c$
and the reduced Planck constant $\hbar$ are equivalent to 1. 
The reduced Planck mass $M_{\rm pl}$ is related to 
the gravitational constant $G$, as $M_{\rm pl}=(8\pi G)^{-1/2}$.
We take present-day Hubble constant 
$H_0=100\,h$\,km~sec$^{-1}$\,Mpc$^{-1}=2.1331 \times 10^{-33}\,h$~eV 
with $h=0.677$, and choose today's density parameters of 
total nonrelativistic matter, baryons, and DE as  
$\Omega_{M0}=0.31$, $\Omega_{b0}=0.05$, and 
$\Omega_{d 0}=0.69$, respectively.
The present-day BEC and CDM density parameters, which are given 
by $\Omega_{\chi 0}$ and $\Omega_{c0}$ respectively, satisfy 
the relation $\Omega_{\chi 0}+\Omega_{c 0}=\Omega_{M0}-\Omega_{b0}=0.26$. 
The scale factor at matter-radiation 
equality is chosen to be $a_{\rm eq}=1/3400$, with today's 
value $a_0=1$.  

\section{Nonrelativistic BEC and background cosmology}
\label{gravitysec}

We begin with a complex scalar field $\chi$ given by the action
\be
{\cal S}=\int {\rm d}^4 x \sqrt{-g} \left[ \frac{R}{16\pi G}
-\nabla^{\mu} \chi^* \nabla_{\mu} \chi-m^2 \chi^* \chi-U(\chi^* \chi) \right]\,,
\label{action}
\ee
where $g$ is the determinant of metric tensor $g_{\mu \nu}$, 
$R$ is the Ricci scalar, and $\nabla^{\mu}$ is the covariant 
derivative operator.
The scalar field has a constant mass $m$ with self-interactions 
described by the potential $U$. 
For an interacting Bose field, the self-coupling potential $U$ 
depends on the particle probability density given by 
\be
\rho=\chi^* \chi\,.
\ee
The two-body interaction is described by the potential 
$U(\rho)=\lambda \rho^2/4$ with a coupling constant $\lambda$.
The attractive and repulsive self-interactions correspond to 
$\lambda<0$ and $\lambda>0$, respectively. 
Many-body interactions contain the terms higher than the order 
$\rho$ in $U(\rho)$.

In this section we only include the field $\chi$ in the matter sector, 
but we will take additional matter sources (CDM, baryons, 
and radiation) into account in Sec.~\ref{persec}. 
Varying the action (\ref{action}) with respect to $\chi^*$, 
it follows that  
\be
\square \chi-m^2 \chi-U_{,\rho}\chi=0\,,
\label{schi}
\ee
where $U_{,\rho} \equiv {\rm d} U/{\rm d} \rho$, and 
\be
\square \chi \equiv g^{\mu \nu} \nabla_{\mu} \nabla_{\nu} \chi
=\frac{1}{\sqrt{-g}} \frac{\partial}{\partial x^{\mu}} 
\left( \sqrt{-g} g^{\mu \nu} \partial_{\nu} \chi \right)\,,
\ee
with the notation $\partial_{\nu}\chi \equiv \partial \chi/\partial x^{\nu}$.

Since the action (\ref{action}) is invariant under a global $U(1)$ 
transformation, there is the current conservation
\be
\nabla^{\mu} j_{\mu}=\frac{1}{\sqrt{-g}} \partial^{\mu} 
\left( \sqrt{-g} j_{\mu} \right)=0\,,
\label{currentcon}
\ee
where 
\be
j_{\mu}=-i \left( \chi^{*} \partial_{\mu} \chi
-\chi \partial_{\mu} \chi^{*} \right)\,.
\label{Jmu}
\ee
Varying the action (\ref{action}) with respect to $g^{\mu \nu}$ 
leads to the Einstein equation 
\be
G_{\mu \nu}=8\pi G\,T_{\mu \nu}\,,
\label{Ein}
\ee
where $G_{\mu \nu}$ is the Einstein tensor, and $T_{\mu \nu}$ 
is the energy-momentum tensor given by 
\be
T_{\mu \nu}=\nabla_{\mu} \chi^* \nabla_{\nu} \chi
+\nabla_{\mu} \chi \nabla_{\nu} \chi^*
-g_{\mu \nu} \left[ g^{\alpha \beta} \nabla_{\alpha} \chi^* 
\nabla_{\beta} \chi+m^2 \chi^* \chi
+U(\rho) \right]\,. 
\label{Tmn}
\ee
We are interested in the cosmology on the spatially flat 
Friedmann-Lema\^itre-Robertson-Walker (FLRW) background 
given by the line element 
\be
{\rm d}s^2=-{\rm d}t^2+a^2(t) \delta_{ij} 
{\rm d}x^i {\rm d}x^j\,,
\label{backmet}
\ee
where $a(t)$ is the time-dependent scale factor.
A key quantity which determines the transition to the BEC formation 
is the Hubble expansion rate $H \equiv \dot{a}/a$ in comparison 
to the mass $m$, where a dot represents the derivative with respect to 
the cosmic time $t$.

\subsection{Covariant hydrodynamical equations}

We derive the hydrodynamical equations of motion in the nonrelativistic 
regime where the field $\chi$ oscillates with the frequency $m$. 
Cosmologically, the field oscillation starts when $H$ drops below 
the order of $m$, more precisely, $H<m/3$ \cite{Marsh:2015xka}.
After averaging over many oscillations, 
the kinetic term $\nabla^{\mu} \chi^*\nabla_{\mu} \chi$ has 
the same contribution to the energy density as $m^2 \chi^* \chi$.
Then, we introduce the energy density associated with 
the massive $\chi$ field, as
\be
\rho_{\chi}=2m^2 \chi^* \chi=2m^2 \rho\,.
\label{rhomdef}
\ee
In the regime $m \gg H$, the $\chi$ field behaves as a single 
classical wave of the nonrelativistic BEC. 
To describe this condensed stage of bosons, 
we take the Madelung representation \cite{Madelung} in the form  
\be
\chi=\sqrt{\frac{\rho_{\chi}}{2m^2}}\,e^{i \theta}\,,
\label{chiMa}
\ee
where the phase part $\theta$ is given by 
\be
\theta=-mt -m v_{\chi}\,.
\ee
The term $-mt$ in $\theta$ characterizes the oscillation 
of $\chi$ induced by the mass $m$.
The scalar quantity $v_{\chi}$, which depends on both time $t$ and 
space $x^i$, corresponds to the velocity potential. 
This latter contribution is dealt as a perturbation 
on the background (\ref{backmet}). 
We note that including the term $-mt$ in $\theta$ allows 
one to eliminate rapidly oscillating terms in $\rho_{\chi}$ 
at the background level. 
The self-coupling term $U_{,\rho}\chi$ in Eq.~(\ref{schi}) leads 
to the deviation from the coherent oscillation of $\chi$ 
with frequency $m$. 
In Sec.~\ref{cossec}, we will derive conditions for the validity 
of the nonrelativistic BEC description in the presence of 
self-couplings.

Substituting Eq.~(\ref{chiMa}) into Eq.~(\ref{Jmu}), 
the current $j_{\mu}$ can be expressed as 
\be
j_{\mu}=\frac{\rho_{\chi}}{m} v_{\mu}=n_{\chi} v_{\mu}\,,
\label{jmuchi}
\ee
where $n_{\chi}=\rho_{\chi}/m$ is the particle number density, 
and $v_{\mu}$ is the four vector field defined by 
\be
v_{\mu} \equiv \frac{\partial_{\mu} \theta}{m}
=\left( -1-\dot{v}_{\chi}, -\partial_i v_{\chi} \right)\,.
\label{vmu}
\ee
{}From the definition (\ref{vmu}), the vector field $v_{\mu}$ 
obeys the irrotational relation 
\be
\nabla_{\mu}v_{\nu}=\nabla_{\nu} v_{\mu}\,.
\label{irro}
\ee
Then, the current conservation (\ref{currentcon}) translates 
to the continuity equation
\be
\nabla^{\mu} \left( \rho_{\chi} v_{\mu} \right)=0\,.
\label{be1}
\ee
Substituting Eq.~(\ref{chiMa}) into Eq.~(\ref{schi}) and 
using Eq.~(\ref{be1}), it follows that \cite{Fagnocchi:2010sn,Bettoni:2013zma}
\be
v^{\mu}v_{\mu}=-1+\beta\,,
\label{vs}
\ee
where 
\be
\beta \equiv \frac{\square \sqrt{\rho_{\chi}}}{m^2 \sqrt{\rho_{\chi}}}
-\frac{U_{,\rho}}{m^2}\,.
\label{betarhom}
\ee
We define the four velocity as  
\be
u_{\mu} \equiv \frac{v_{\mu}}{\sqrt{-v_\alpha v^{\alpha}}}\,,
\ee
which satisfies the normalization $u_{\mu} u^{\mu}=-1$.
Taking the covariant derivative of Eq.~(\ref{vs}) and exploiting
the irrotational property (\ref{irro}) of $v_{\mu}$, we obtain
\be
v^{\mu} \nabla_{\mu} v_{\nu}=\frac{1}{2} \nabla_{\nu} \beta\,.
\label{be2}
\ee
This is the analogue of the Euler equation in the hydrodynamical mechanics.

Substituting Eq.~(\ref{chiMa}) into Eq.~(\ref{Tmn}), 
the matter energy-momentum tensor can be expressed as 
\be
T_{\mu \nu}=\rho_{\chi} v_{\mu} v_{\nu}+\frac{\nabla_{\mu}\rho_{\chi}
\nabla_{\nu}\rho_{\chi}}{4m^2 \rho_{\chi}}
-g_{\mu \nu} \left[ \frac{\rho_{\chi}}{2} \left( v^{\alpha}v_{\alpha}
+1 \right)+\frac{\nabla^{\alpha} \rho_{\chi} \nabla_{\alpha} 
\rho_{\chi}} {8 m^2 \rho_{\chi}}+U(\rho) \right]\,.
\label{Tmnd}
\ee
The dynamics of nonrelativistic BEC 
is governed by the continuity Eq.~(\ref{be1}) and Euler Eq.~(\ref{be2}) 
as well as by the Einstein Eq.~(\ref{Ein}) with the 
energy-momentum tensor (\ref{Tmnd}).

\subsection{FLRW background}
\label{cossec}

Let us consider the flat FLRW background given by the 
line element (\ref{backmet}).
Since $v_{\chi}=0$ on this background, the vector field 
$v_{\mu}$ in Eq.~(\ref{vmu}) reduces to 
the four velocity $u_{\mu}=(-1,0,0,0)$. 
Then, the continuity Eq.~(\ref{be1}) gives
\be
\dot{\rho}_{\chi}+3H \rho_{\chi}=0\,,
\label{coneq2}
\ee
so that the BEC energy density decreases 
as $\rho_{\chi} \propto a^{-3}$.
{}From Eq.~(\ref{vs}) we have 
\be
\beta=0\,,
\label{beta0}
\ee
where
\be
\beta=\frac{\dot{\rho}_{\chi}^2-2\rho_{\chi} (\ddot{\rho}_{\chi}
+3H \dot{\rho}_{\chi})}{4m^2 \rho_{\chi}^2}-\frac{U_{,\rho}}{m^2}
=\frac{3(2\dot{H}+3H^2)}{4m^2}-\frac{U_{,\rho}}{m^2}\,.
\label{betaeq}
\ee
In the second equality of Eq.~(\ref{betaeq}), 
we used Eq.~(\ref{coneq2}) and its time derivative.
During exact matter dominance ($a \propto t^{2/3}$)
there is the relation $2\dot{H}+3H^2=0$, in which case
the property (\ref{beta0}) holds in the absence of 
the self-coupling potential ($U=0$). 
During the radiation era ($a \propto t^{1/2}$), 
we have $\dot{H}+2H^2=0$ and hence
$\beta=-3H^2/(4m^2)-U_{,\rho}/m^2$.
Then, the relation (\ref{beta0}) approximately holds for
\be
m^2 \gg H^2\,,\quad {\rm and} \quad 
m^2 \gg \left| U_{,\rho} \right|\,.
\label{mcon}
\ee
These conditions ensure the validity of the 
nonrelativitistic BEC description 
based on the Madelung representation (\ref{chiMa}).
We note that the Euler Eq.~(\ref{be2}) is trivially satisfied 
on the FLRW spacetime.

On the background (\ref{backmet}), the scalar-field equation 
(\ref{schi}) yields
\be
\ddot{\chi}+3H \dot{\chi}
+(m^2+U_{,\rho})\chi=0\,.
\label{ttchi}
\ee
Under the conditions (\ref{mcon}), Eq.~(\ref{ttchi}) 
approximately yields $(a^{3/2} \chi)^{\cdot\cdot}
+m^2 (a^{3/2} \chi) \simeq 0$. 
Then, the scalar field exhibits a damped oscillation described by 
the solution $\chi=\chi_i (a/a_i)^{-3/2} e^{-imt}$, where 
$\chi_i$ and $a_i$ are constants.
Comparing the amplitude of this solution with 
Eq.~(\ref{chiMa}), the field energy density evolves as 
$\rho_{\chi}=2m^2 \chi_i^2 (a/a_i)^{-3}$, which is consistent 
with Eq.~(\ref{coneq2}). 
In the early cosmological epoch where the condition $m^2 \ll H^2$ 
is satisfied, the field slowly evolves along the potential 
with the second time derivative $\ddot{\chi}$ negligible relative to 
the other terms in Eq.~(\ref{ttchi}). 
After $H$ drops below the order of $m$, the scalar field 
starts to oscillate around the potential minimum. The 
onset of this oscillation is characterized by 
the condition $m \gtrsim 3H$ \cite{Hlozek:2014lca,Marsh:2015xka}.

Substituting Eq.~(\ref{vmu}) into Eq.~(\ref{Tmnd}) and using the relation 
(\ref{beta0}) with Eq.~(\ref{betaeq}), the nonvanishing components of $T_{\mu \nu}$ 
are $T_{00}=\rho_{\rm eff}$ and $T_{ij}=a^2 P_{\rm eff} \delta_{ij}$, 
where $\rho_{\rm eff}$ and $P_{\rm eff} $ are the effective 
field energy density and pressure given by 
\be
\rho_{\rm eff} 
= \rho_{\chi}+\frac{\dot{\rho}_{\chi}^2}{8m^2 \rho_{\chi}}+U\,,
\qquad
P_{\rm eff} = 
\frac{\dot{\rho}_{\chi}^2}{8m^2 \rho_{\chi}}-U\,.
\label{rhoPeff}
\ee
{}From the Einstein equation (\ref{Ein}), we obtain
\ba
& &
3H^2=8 \pi G \rho_{\rm eff}\,,
\label{back1}\\
& &
3H^2+2\dot{H}=-8 \pi G P_{\rm eff}\,.
\label{back2}
\ea
Under the first condition of Eq.~(\ref{mcon}),
the term $\dot{\rho}_{\chi}^2/(8m^2 \rho_{\chi})$ 
in Eq.~(\ref{rhoPeff}), which is identical to 
$9H^2 \rho_{\chi}/(8m^2)$, is suppressed 
relative to $\rho_{\chi}$. 
As long as the condition 
\be
\rho_{\chi} \gg |U|
\label{mcon2}
\ee
is satisfied, it follows that $\rho_{\rm eff} \simeq \rho_{\chi}$ 
and $|P_{\rm eff}/\rho_{\rm eff}| \ll 1$. 
In this case, the cosmological dynamics dominated by 
the rapidly oscillating scalar field over the Hubble time scale 
$H^{-1}$ is equivalent to that of the matter era characterized 
by $\dot{H} \simeq -3H^2/2$ and $a \propto t^{2/3}$. 

The above result shows that, under the conditions (\ref{mcon}) 
and (\ref{mcon2}), the oscillating scalar field can be the source 
for DM in a state of the nonrelativistic BEC. 
Using the subscript ``osc'' at the onset of oscillations, 
the field energy density today ($a=1$) is given by 
$\rho_{\chi 0}=(\rho_{\chi})_{\rm osc}a_{\rm osc}^3$. 
By the end of this section, we consider the case in which 
the quadratic potential $m^2 |\chi|^2$ dominates 
over $U$. Then, the initial field density can be estimated as 
$(\rho_{\chi})_{\rm osc} \simeq m^2 |\chi_{\rm osc}|^2$.
Then, today's density parameter of the field $\chi$ is given by 
\be
\Omega_{\chi 0}=\frac{\rho_{\chi 0}}
{3M_{\rm pl}^2 H_0^2} 
\simeq \frac{m^2 |\chi_{\rm osc}|^2 a_{\rm osc}^3}{3M_{\rm pl}^2 H_0^2}\,,
\label{Om0}
\ee
where $M_{\rm pl}=(8\pi G)^{-1/2}$ is the reduced Planck mass 
and $H_0=2.1331 \times 10^{-33}\,h$~eV is the Hubble constant.
To estimate the Hubble expansion rate as a function of the scale factor, 
we take radiation, CDM, baryons, and DE into account. 
Expressing today's density parameters of total 
nonrelativistic matter and DE as $\Omega_{M0}$ and 
$\Omega_{d 0}$, respectively, and assuming that the origin 
of DE is the cosmological constant, the Hubble parameter 
can be expressed as
\be
H(a)=H_0 \sqrt{\Omega_{M0}(a+a_{\rm eq}) a^{-4} 
+\Omega_{d 0}}\,,
\label{Ha}
\ee
where $a_{\rm eq}=\Omega_{r0}/\Omega_{M0}\simeq 1/3400$
is the scale factor at matter-radiation equality 
($\Omega_{r0}$ is today's radiation density parameter).
The scale factor $a_*$ at the onset of oscillations can be 
identified by the condition 
\be
m=3H(a_{\rm osc})\,.
\label{mH}
\ee
The Hubble parameter at matter-radiation equality is given by 
$H(a_{\rm eq}) \simeq H_0 \sqrt{2 \Omega_{M0}a_{\rm eq}^{-3}}
\simeq 2.3 \times 10^{-28}$~eV, where we used the 
values $\Omega_{M0}=0.31$ and $h=0.677$. 
For $m>7 \times 10^{-28}$~eV the scalar field starts to oscillate 
during radiation domination, while, for $m<7 \times 10^{-28}$~eV, 
the oscillation begins in the matter era.

During radiation dominance ($a \ll a_{\rm eq}$) the 
Hubble parameter is approximately given by 
$H(a) \simeq H_0 \sqrt{\Omega_{r0} a^{-4}}$, so that 
$a_{\rm osc} \simeq (9H_0^2 \Omega_{r0}/m^2)^{1/4}$. 
Substituting this relation into Eq.~(\ref{Om0}), 
it follows that \cite{Hlozek:2014lca,Marsh:2015xka}
\be
\Omega_{\chi0} \simeq \sqrt{3} \Omega_{r0}^{3/4} 
\left( \frac{m}{H_0} \right)^{1/2} 
\left( \frac{|\chi_{\rm osc}|}{M_{\rm pl}} \right)^2 \qquad 
{\rm for}\quad m>7 \times 10^{-28}~{\rm eV}.
\ee
During matter dominance we have 
$H(a) \simeq H_0 \sqrt{\Omega_{M0} a^{-3}}$ and hence 
$a_{\rm osc} \simeq (9H_0^2 \Omega_{M0}/m^2)^{1/3}$. 
Then, Eq.~(\ref{Om0}) reduces to 
\be
\Omega_{\chi 0} \simeq 3 \Omega_{M0} 
\left( \frac{|\chi_{\rm osc}|}{M_{\rm pl}} \right)^{2} \qquad 
{\rm for}\quad 1 \times 10^{-33}~{\rm eV}  \ll m<7 \times 10^{-28}~{\rm eV}.
\ee
For $m={\cal O}(10^{-33})$~eV, the field is nearly frozen until recently, 
but this is the region in which the field energy density works as 
DE rather than DM.

\section{Cosmological perturbations}
\label{persec}

Now, we proceed to the study of linear cosmological perturbations 
on top of the flat FLRW background. 
Besides the $\chi$ field described by the action 
(\ref{action}), we take the perfect fluids of 
baryons, CDM, DE, and radiation (photons and neutrinos) 
into account, which are labelled by $b$, $c$, $d$, and $r$ ($\gamma$ and $\nu$)
respectively. If the $\chi$ field is responsible for all DM,
we do not need to include CDM in the matter action. 
The source for DE can be the cosmological constant or 
other dynamical fields \cite{Copeland:2006wr}, but it is also possible to realize 
the late-time cosmic acceleration in DE models of 
a perfect fluid \cite{Jimenez:2020npm}. 
Then, the total system is described by the action 
\be
{\cal S}=
\int {\rm d}^4 x \sqrt{-g} \left[ \frac{R}{16\pi G}
-\nabla^{\mu} \chi^* \nabla_{\mu} \chi-m^2 \chi^* \chi-U(\chi^* \chi) \right]
+\int {\rm d}^{4}x\,L_{\rm pf}\,,
\label{SMtotal}
\ee
where $L_{\rm pf}$ is the perfect-fluid Lagrangian 
given by \cite{Sorkin,Brown,DGS,Amendola:2020ldb,Kase:2020hst}
\be
L_{\rm pf}=-\sum_{I=b,c,d,r} \sqrt{-g} \left[ \rho_I(n_I)
+j_I^{\mu} \partial_{\mu} \ell_I \right]\,.
\label{LM}
\ee
The Lagrangian (\ref{LM}) consists of the energy density 
$\rho_I$, the current vector $j_I^{\mu}$, and the Lagrange multiplier 
$\ell_I$, where $\rho_I$ is a function of the fluid number density $n_I$.
Varying the Lagrangian $L_{\rm pf}$ with respect to $\ell_I$, there is 
the current conservation\footnote{In Refs.~\cite{Jimenez:2020npm,Amendola:2020ldb,Kase:2020hst}
the quantity $J_I^{\mu}=\sqrt{-g}\,j_I^{\mu}$ is used instead of $j_I^{\mu}$, in which 
case the current conservation (\ref{Jmucon}) is expressed as $\partial_{\mu}J_I^{\mu}=0$.} 
\be
\nabla_{\mu} j_I^{\mu}=\frac{1}{\sqrt{-g}}
\partial_{\mu}  \left( \sqrt{-g}\,j_I^{\mu} \right)=0\,,
\label{Jmucon}
\ee
which is analogous to Eq.~(\ref{currentcon}) of the complex 
scalar field $\chi$. The number density and four velocity 
of each matter species are given, respectively, by 
\ba
n_I &=& \sqrt{-g_{\mu \nu} j_I^{\mu} j_I^{\nu}}\,,
\label{nI0}\\
u_{I \mu} &=& \frac{j_{I \mu}}{n_I}\,.
\label{nI}
\ea
{}From Eqs.~(\ref{nI0}) and (\ref{nI}), the four velocity 
$u_{I \mu}$ satisfies 
the normalization
\be
u_{I \mu} u_I^{\mu}=-1\,.
\label{unor}
\ee
The variation of $L_{\rm pf}$ with respect to $j_I^{\mu}$ 
leads to 
\be
\partial_{\mu} \ell_I=\rho_{I,n_I}u_{I \mu}\,.
\label{pll}
\ee
On using the relation (\ref{nI}), the current conservation 
(\ref{Jmucon}) translates to 
\be
\nabla_{\mu} \left( n_I u_I^{\mu} \right)=0\,,
\label{coneq}
\ee
or equivalently, 
\be
u_I^{\mu} \nabla_{\mu} \rho_I+ \left( \rho_I+P_I \right) 
\nabla_{\mu} u_I^{\mu}=0\,,
\label{ucon}
\ee
where $P_I$ is the pressure defined by 
\be
P_I \equiv n_I \rho_{I,n_I}-\rho_I\,.
\label{PI}
\ee
For nonrelativistic matter with $n_I=\rho_I/m_I$ and mass $m_I$, 
the continuity Eq.~(\ref{coneq}) is analogous to Eq.~(\ref{be1}) of the BEC. 
We note, however, that the four vector $v^{\mu}$ does not 
correspond to the four velocity $u_I^{\mu}$, so there is the 
difference between Eqs.~(\ref{be1}) and (\ref{coneq}) 
at the level of perturbations (as we will see in Sec.~\ref{greadysec}).

On using Eq.~(\ref{pll}) with Eqs.~(\ref{nI}) and (\ref{PI}), 
the current vector $j_{I \mu}$ is related to the Lagrange 
multipler $\ell_I$ as 
\be
j_{I \mu}=\frac{n_I^2}{\rho_I+P_I} 
\partial_{\mu} \ell_I\,. 
\label{jim}
\ee
On the other hand, the current (\ref{jmuchi}) of the BEC 
is given by  $j_{\mu}=(\rho_{\chi}/m^2) \partial_{\mu} \theta$, 
where we used Eq.~(\ref{vmu}).
Since there is the relation $m=\rho_{\chi}/n_{\chi}$ for the 
nonrelativistic BEC, this current reduces to $j_{\mu}=(n_{\chi}^2/\rho_{\chi}) 
\partial_{\mu} \theta$. 
Comparing it to Eq.~(\ref{jim}), the quantity $\theta$ 
in the Madelung representation 
(\ref{chiMa}) has the correspondence with 
$\ell_I$ in the perfect-fluid Lagrangian (\ref{LM}) with 
the vanishing pressure ($P_I=0$).

Taking the covariant derivative of Eq.~(\ref{pll}), the four 
velocity $u_{I \mu}$ satisfies the relation 
\be
\nabla_{\nu} u_{I \mu}-\nabla_{\mu} u_{I \nu}
=\frac{1}{\rho_{I,n_I}} \left( u_{I \nu} \nabla_{\mu} \rho_{I,n_I}
-u_{I \mu} \nabla_{\nu} \rho_{I,n_I} \right)\,.
\label{nabmu}
\ee
If we consider nonrelativistic matter with the mass $m_I$ and 
density $\rho_I=m_I n_I$, we have $\rho_{I,n_I}=m_I={\rm constant}$ 
and hence $\nabla_{\nu} u_{I \mu}=\nabla_{\mu} u_{I \nu}$. 
Exerting the operator $\nabla_{\nu}$ for Eq.~(\ref{unor}) 
and using the property (\ref{nabmu}), it follows that  
\be
u_I^{\mu} \nabla_{\mu} u_{I \nu}
=-\frac{1}{\rho_{I,n_I}} \left( u_{I\nu} u_I^{\mu} 
\nabla_{\mu} \rho_{I,n_I}+\nabla_{\nu} \rho_{I,n_I} 
\right)\,,
\label{Euu}
\ee
which corresponds to the Euler equation for the perfect fluid. 
For nonrelativistic matter the right hand-side 
of Eq.~(\ref{Euu}) vanishes, so that  the Euler equation 
is simplified to $u_I^{\mu} \nabla_{\mu} u_{I \nu}=0$.
The Euler Eq.~(\ref{be2}) of the BEC 
is different from that of the nonrelativistic perfect fluid, 
in that the right hand-side of Eq.~(\ref{be2}) contains 
the derivative term $\nabla_{\nu}\beta/2$.

The energy-momentum tensor $T_{\mu \nu}^{({\rm pf})}$ associated 
with the perfect-fluid Lagrangian (\ref{LM}) follows by its variation 
with respect to $g^{\mu \nu}$. 
On using the properties 
$\delta \sqrt{-g}/\delta g^{\mu \nu}=-(1/2)\sqrt{-g} g_{\mu \nu}$, 
$\delta n_I/\delta g^{\mu \nu}=(n_I/2)(g_{\mu \nu}-u_{I \mu}u_{I \nu})$, 
and Eq.~(\ref{pll}), we have
\be
T_{\mu \nu}^{({\rm pf})}= -\frac{2}{\sqrt{-g}} 
\frac{\delta L_{\rm pf}}{\delta g^{\mu \nu}}
=\sum_{I} \left[ \left( \rho_I+P_I \right) 
u_{I \mu} u_{I \nu}+P_I g_{\mu \nu} \right]\,.
\label{Tmnp}
\ee
Varying the total action (\ref{SMtotal}) with respect to 
$g^{\mu \nu}$, we obtain the gravitational field 
equation of motion 
\be
G_{\mu \nu}=8\pi G \left[ T_{\mu \nu}+T_{\mu \nu}^{(\rm pf)} 
\right]\,,
\label{Einf}
\ee
where $T_{\mu \nu}$ is given by Eq.~(\ref{Tmn}). 
Since we are interested in the regime where the BEC is formed, 
we will employ the energy-momentum tensor of the form (\ref{Tmnd}) 
in the following.

\subsection{Perturbation equations in a gauge-ready form}
\label{greadysec}

The general perturbed line element containing four scalar 
metric perturbations  is given 
by \cite{Bardeen:1980kt,Kodama:1985bj,Mukhanov:1990me}
\be
{\rm d}s^2=-(1+2\alpha) {\rm d}t^2
+2 \partial_i B {\rm d}t {\rm d}x^i
+a^2(t) \left[ (1+2\zeta) \delta_{ij}
+2\partial_i \partial_j E \right] {\rm d}x^i {\rm d}x^j\,,
\label{permet}
\ee
where $\alpha, B, \zeta, E$ depend on both cosmic time $t$ 
and spatial coordinates $x^i$. 
Unlike Refs.~\cite{DeFelice:2016yws,Heisenberg:2018mxx}, 
we do not take intrinsic vector perturbations into account 
as they are nondynamical 
for the theory under consideration.
The evolution of tensor perturbations is the same as that in 
standard general relativity. 
We first derive the linear perturbation equations 
of motion without choosing particular gauges and then 
express them in terms of gauge-invariant quantities.

For the nonrelativistic BEC, the energy 
density $\rho_{\chi}$ and the quantity $\beta$, which are defined 
respectively by Eqs.~(\ref{rhomdef}) and (\ref{betarhom}), are 
decomposed into the background and perturbed parts, as 
\be
\rho_{\chi}=\bar{\rho}_\chi+\delta \rho_\chi\,,\qquad 
\beta=\bar{\beta}+\delta \beta\,,
\label{rhobeta}
\ee
where a bar represents the background values, 
and $\bar{\beta}=0$. 
As we will see below, the perturbation equations for matter perturbation 
$\delta \rho_\chi$ and velocity potential $v_\chi$ follow 
by substituting Eqs.~(\ref{vmu}) and (\ref{rhobeta}) into the continuity 
Eq.~(\ref{be1}) and the Euler Eq.~(\ref{be2}).

For the perfect-fluid sector, the fluid number density (\ref{nI0}) is decomposed into the 
background and perturbed parts, as $n_I=\bar{n}_I +\delta n_I$.
Since the fluid density $\rho_I$ depends on its number density 
$n_I$, the matter perturbation is given by 
\be
\delta \rho_I=\rho_{I,n_I} \delta n_I
=\frac{\rho_I+P_I}{n_I} \delta n_I\,.
\ee
Here and in the following, we omit a bar from the background quantities.
For the line element (\ref{permet}), the temporal and 
spatial components of the fluid four velocity $u_{I \mu}$, 
up to first order in perturbations, are \cite{Amendola:2020ldb,Kase:2020hst} 
\be
u_{I0}=-1-\alpha\,,\qquad
u_{Ii}=-\partial_i v_I\,, 
\label{uI}
\ee
where $v_I$ is the velocity potential.
{}From Eq.~(\ref{nI}), the components of $j_{I \mu}$ are expressed as 
\be
j_{I 0}=-n_I-n_I \alpha-\frac{n_I}{\rho_I+P_I} 
\delta \rho_I\,,\qquad 
j_{Ii}=-n_I \partial_i v_I\,.
\ee
In the following, we will use $\delta \rho_I$ and $v_I$ 
instead of $j_{I 0}$ and $j_{Ii}$ for the derivation of 
perturbation equations of motion.
We also introduce the equation of state and 
adiabatic sound speed squared, as
\be
w_I=\frac{P_I}{\rho_I}\,,\qquad
c_I^2=\frac{\dot{P}_I}{\dot{\rho}_I}
=\frac{n_I \rho_{I,n_I n_I}}{\rho_{I,n_I}}\,.
\label{cI}
\ee

Let us first consider the background equations of motion. 
For the nonrelativistic BEC, the $\chi$ field obeys the continuity 
equation (\ref{coneq2}), i.e., 
\be
\dot{\rho}_{\chi}+3H \rho_{\chi}=0\,.
\label{backcon1}
\ee
On the flat FLRW background (\ref{backmet}), the four velocity 
of each perfect fluid is given by $u_{I \mu}=(-1,0,0,0)$,
so the continuity Eq.~(\ref{ucon}) leads to
\be
\dot{\rho}_I+3H \left( 1+w_I \right) \rho_I=0\,,\qquad 
{\rm for} \quad I=b,c,d,r\,,
\label{backcon2}
\ee
whereas the Euler Eq.~(\ref{Euu}) trivially holds.
{}From Eq.~(\ref{Tmnp}), the nonvanishing components of perfect-fluid
energy-momentum tensors are given by 
$T_{00}=\sum_{I} \rho_I$ and $T_{ij}=\sum_{I} a^2 P_{I} \delta_{ij}$. 
Then, the $(00)$ and $(ii)$ components of the Einstein 
Eq.~(\ref{Einf}) give
\ba
& &
3H^2=8 \pi G \biggl( \rho_{\chi}
+\frac{\dot{\rho}_{\chi}^2}{8m^2 \rho_{\chi}}+U+\sum_{I} 
\rho_I \biggr)\,,
\label{back1f}\\
& &
3H^2+2\dot{H}=-8 \pi G \biggl( \frac{\dot{\rho}_{\chi}^2}
{8m^2 \rho_{\chi}}-U
+\sum_{I} w_I \rho_I \biggr)\,,
\label{back2f}
\ea
respectively, where we used Eq.~(\ref{rhoPeff}).

Expanding Eqs.~(\ref{be1}) and (\ref{be2}) up to first 
order in perturbations, the matter perturbation 
$\delta \rho_{\chi}$ and velocity potential $v_{\chi}$ obey 
\ba
& &
\dot{\delta \rho}_{\chi}+3H \delta \rho_{\chi}
+\rho_{\chi} \left( 3\dot{\zeta}
-\frac{\partial^2 v_{\chi}}{a^2} 
-\frac{\partial^2 B}{a^2}
+\partial^2 \dot{E} \right)
-\frac{1}{2} \rho_{\chi}  \dot{\delta \beta}=0\,,
\label{pereq1}\\
& &
\dot{v}_{\chi}-\alpha+\frac{1}{2} \delta \beta=0\,,
\label{pereq2}
\ea
where $\partial^2 \equiv \sum_{i=1}^3 \partial_i^2$, and 
\ba
\delta \beta
&=& \frac{1}{2m^2 \rho_{\chi}} \biggl[ 
\frac{1}{a^2} \left(  
\partial^2 \delta \rho_{\chi}+\dot{\rho}_{\chi} \partial^2 
B \right)+2\ddot{\rho}_{\chi} \alpha+\dot{\rho}_{\chi} 
\left( 6H \alpha+\dot{\alpha} 
-3\dot{\zeta}-\partial^2 \dot{E} \right)
-\ddot{\delta \rho}_{\chi}-3H \dot{\delta \rho}_{\chi}
-\frac{\dot{\rho}_{\chi}^2}{\rho_{\chi}^2}\delta \rho_{\chi} \nonumber \\
&&\qquad \qquad
+\frac{1}{\rho_{\chi}} \left\{ \ddot{\rho}_{\chi} \delta \rho_{\chi}
+\dot{\rho}_{\chi} \left( \dot{\delta \rho}_{\chi}
+3H \delta \rho_{\chi} \right)
-\dot{\rho}_{\chi}^2 \alpha \right\}
-\frac{\rho_{\chi} U_{,\rho \rho}}{m^2}\delta \rho_{\chi} \biggr]\,.
\label{dbeta}
\ea
The linearly perturbed continuity and Euler Eqs.~(\ref{ucon}) and (\ref{Euu}) 
for perfect fluids are given, respectively, by 
\ba
& &
\dot{\delta \rho}_I+3H \left( 1+c_I^2 \right) \delta \rho_I
+\rho_I \left( 1+w_I \right) \left( 3\dot{\zeta}
-\frac{\partial^2 v_I}{a^2} 
-\frac{\partial^2 B}{a^2}
+\partial^2 \dot{E} \right)=0\,,
\label{pereq3}\\
& &
\dot{v}_I-3H c_I^2 v_I-\alpha-c_I^2 \frac{\delta \rho_I}
{\rho_I (1+w_I)}=0\,,
\label{pereq4}
\ea
which hold for each $I=b,c,d,r$.

The energy-momentum tensors of the BEC and perfect fluids 
are given, respectively, by Eqs.~(\ref{Tmnd}) and (\ref{Tmnp}). 
{}From the (00), $(0i)$, trace, and traceless components of 
the perturbed Einstein Eq.~(\ref{Einf}), we obtain
\ba
\hspace{-0.5cm}
& & 
6H \left( H \alpha -\dot{\zeta} \right)
+2 \left( \frac{\partial^2 \zeta}{a^2}+\frac{H}{a^2}\partial^2 B
-H \partial^2 \dot{E} \right) \nonumber \\
\hspace{-0.5cm}
& &
+8\pi G \biggl( \delta \rho_{\chi}-\frac{9H^2-4U_{,\rho}}{8m^2} 
\delta \rho_{\chi}
-\frac{3H \dot{\delta \rho}_{\chi}}{4m^2}
-\frac{9H^2}{4m^2} \rho_{\chi} \alpha-\frac{1}{2}\rho_{\chi} 
\delta \beta+\sum_{I}\delta \rho_I \biggr)=0\,,
\label{pereq5}\\
\hspace{-0.5cm}
& &
H \alpha-\dot{\zeta}-4\pi G \biggl[ \rho_{\chi} v_{\chi}
-\frac{3H}{4m^2} \delta \rho_{\chi}
+\sum_{I} \rho_I \left( 1+w_I \right) v_I \biggr]=0\,,
\label{pereq6}\\
\hspace{-0.5cm}
& &
\ddot{\zeta}+3H \dot{\zeta}-H \dot{\alpha} 
-\left( 3H^2+2\dot{H} \right) \alpha
-4\pi G \biggl( \frac{9H^2+4U_{,\rho}}{8m^2} 
\delta \rho_{\chi}
+\frac{3H \dot{\delta \rho}_{\chi}}{4m^2}
+\frac{9H^2}{4m^2} \rho_{\chi} \alpha
+\frac{1}{2} \rho_{\chi} \delta \beta
-\sum_{I}c_I^2 \delta \rho_I \biggr)=0\,,
\label{pereq7}\\
\hspace{-0.5cm}
& &
\alpha+\zeta+\dot{B}+HB
-a^2 \left( \ddot{E}+3H \dot{E} \right)=0\,.
\label{pereq8}
\ea
The perturbation Eqs.~(\ref{pereq1})-(\ref{pereq8}) are written 
in a gauge-ready form \cite{Hwang,Heisenberg:2018wye}, 
i.e., they are ready for fixing any gauge conditions.

\subsection{Gauge-invariant perturbation equations}

To study the evolution of cosmological perturbations without worrying 
about unphysical gauge degrees of freedom, we consider
gauge-invariant perturbations invariant under the infinitesimal 
coordinate transformation 
$t \to t+\xi^{0}$ and $x^{i} \to x^{i}+\delta^{ij} \partial_{j} \xi$.
We introduce the following gauge-invariant 
combinations \cite{Bardeen:1980kt} 
\ba
& &
\Psi=\alpha+\frac{{\rm d}}{{\rm d}t} 
\left( B - a^2 \dot{E} \right)\,,\qquad 
\Phi=-\zeta+H \left( B - a^2 \dot{E} \right)\,,
\nonumber \\
& &
\delta\rho_{\chi \scriptsize{{\rm N}}}=
\delta \rho_{\chi}+\dot{\rho}_{\chi} 
\left( B-a^2 \dot{E}\right)\,,
\qquad
\delta\rho_{I{\rm N}}=\delta \rho_I+\dot{\rho}_I 
\left( B-a^2 \dot{E}\right)\,,
\nonumber \\
& & 
v_{\chi {\rm N}}=v_{\chi}+B-a^2 \dot{E}\,,\qquad
v_{I{\rm N}}=v_I+B-a^2 \dot{E}\,.
\label{delphiN}
\ea
It is possible to express Eqs.~(\ref{pereq1})-(\ref{pereq2}), 
(\ref{pereq3})-(\ref{pereq4}), and (\ref{pereq5})-(\ref{pereq8})
in terms of the above gauge-invariant variables. 
On using the background equations of motion, 
all the gauge-dependent quantities such as $B$ and $E$ 
disappear from the perturbation equations, so that  
\ba
\hspace{-0.8cm}
& &
\dot{\delta \rho}_{\chi{\rm N}}+3H \delta \rho_{\chi {\rm N}}
-\rho_{\chi} \left( 3\dot{\Phi}
+\frac{\partial^2 v_{\chi{\rm N}}}{a^2} \right)
-\frac{1}{2} \rho_{\chi} \dot{\delta \beta_{\rm N}}=0\,, 
\label{pereq1d}\\
\hspace{-0.8cm}
& &
\dot{v}_{\chi{\rm N}}-\Psi+\frac{1}{2} \delta \beta_{\rm N}=0\,,
\label{pereq2d}\\
\hspace{-0.8cm}
& &
\dot{\delta \rho}_{I{\rm N}}+3H \left( 1+c_I^2 \right) \delta \rho_{I{\rm N}}
-\rho_I \left( 1+w_I \right) \left( 3\dot{\Phi}
+\frac{\partial^2 v_{I{\rm N}}}{a^2} \right)=0\,,
\label{pereq3d}\\
\hspace{-0.8cm}
& &
\dot{v}_{I{\rm N}}-3H c_I^2 v_{I{\rm N}}-\Psi-c_I^2 \frac{\delta \rho_{I{\rm N}}}
{\rho_I (1+w_I)}=0\,,
\label{pereq4d}\\
\hspace{-0.8cm}
& & 
6H \left( H \Psi +\dot{\Phi} \right)
-2 \frac{\partial^2 \Phi}{a^2}
+8\pi G \biggl( \delta \rho_{\chi{\rm N}}
-\frac{9H^2-4U_{,\rho}}{8m^2} 
\delta \rho_{\chi{\rm N}}
-\frac{3H \dot{\delta \rho}_{\chi{\rm N}}}{4m^2}
-\frac{9H^2}{4m^2} \rho_\chi \Psi-\frac{1}{2}\rho_\chi 
\delta \beta_{\rm N}
+\sum_{I}\delta \rho_{I{\rm N}} \biggr)=0\,,
\label{pereq5d}\\
\hspace{-0.8cm}
& &
H \Psi+\dot{\Phi}-4\pi G \biggl[ \rho_\chi v_{\chi{\rm N}}
-\frac{3H}{4m^2} \delta \rho_{\chi{\rm N}}
+\sum_{I} \rho_I \left( 1+w_I \right) v_{I{\rm N}}  \biggr]=0\,,
\label{pereq6d}\\
\hspace{-0.8cm}
& &
\ddot{\Phi}+3H \dot{\Phi}+H \dot{\Psi} 
+\left( 3H^2+2\dot{H} \right) \Psi \nonumber \\
\hspace{-0.8cm}
& &
+4\pi G \biggl( \frac{9H^2+4U_{,\rho}}{8m^2} 
\delta \rho_{\chi{\rm N}}
+\frac{3H \dot{\delta \rho}_{\chi{\rm N}}}{4m^2}
+\frac{9H^2}{4m^2} \rho_\chi \Psi
+\frac{1}{2} \rho_\chi \delta \beta_{\rm N}
-\sum_{I}c_I^2 \delta \rho_{I{\rm N}} \biggr)=0\,,
\label{pereq7d}\\
\hspace{-0.8cm}
& &
\Psi=\Phi\,,
\label{pereq8d}
\ea
where $\delta \beta_{\rm N}$ is the gauge-invariant variable 
given by 
\ba
\delta \beta_{\rm N} &=& \delta \beta+\dot{\beta} 
\left( B-a^2 \dot{E} \right) \nonumber \\
&=& \frac{1}{2m^2 \rho_\chi} \biggl[ 
\frac{1}{a^2}
\partial^2 \delta \rho_{\chi{\rm N}}
+2\ddot{\rho}_\chi \Psi+\dot{\rho}_\chi 
\left( 6H \Psi+\dot{\Psi} +3\dot{\Phi} \right)
-\ddot{\delta \rho}_{\chi{\rm N}}-3H \dot{\delta \rho}_{{\chi\rm N}}
-\frac{\dot{\rho}_\chi^2}{\rho_\chi^2} \delta \rho_{\chi{\rm N}} \nonumber \\
&&\qquad \qquad
+\frac{1}{\rho_\chi} \left\{ \ddot{\rho}_\chi \delta \rho_{\chi{\rm N}}
+\dot{\rho}_\chi \left( \dot{\delta \rho}_{\chi{\rm N}}
+3H \delta \rho_{\chi{\rm N}} \right)
-\dot{\rho}_\chi^2 \Psi \right\}
-\frac{\rho_{\chi} U_{,\rho \rho}}{m^2}\delta \rho_{\chi{\rm N}} \biggr]\,.
\label{delbeN}
\ea
Since the background value of $\beta$ vanishes, $\delta \beta_{\rm N}$ 
is identical to $\delta \beta$. 

After the CMB recombination epoch, the equations of state  
and the sound speed squares for CDM and baryons can be 
taken to be $w_c=w_b=0$ and $c_c^2=c_b^2=0$, while the photons and 
relativistic neutrinos have the values $w_\gamma=w_\nu=1/3$ and 
$c_\gamma^2=c_\nu^2=1/3$. The DE equation of state 
$w_d$ needs to be close to $-1$ at low redshifts, while 
the sound speed squared $c_d^2$ should not be much smaller than 1
to avoid the clustering of DE perturbations \cite{Jimenez:2020npm}.

Prior to the recombination, the baryons and photons are tightly 
coupled to each other due to the Thomson scattering weighed 
by the product of cross section $\sigma_{\rm T}$ and 
electron number density $n_{\rm e}$. 
Taking into account this coupling, the perturbation 
Eq.~(\ref{pereq4d}) of velocity potentials for baryons and photons 
are modified, respectively, to \cite{Dodelson}
\ba
& &
\dot{v}_{b{\rm N}}-\Psi=-\frac{4\rho_{\gamma}}{3\rho_b} 
\sigma_{\rm T} n_{\rm e} \left( v_{b{\rm N}}-v_{\gamma {\rm N}} 
\right)\,,\label{vbeq}\\
& & 
\dot{v}_{\gamma{\rm N}}-H v_{\gamma{\rm N}}-\Psi
-\frac{1}{4}\frac{\delta \rho_{\gamma{\rm N}}}
{\rho_\gamma}=\sigma_{\rm T} n_{\rm e} \left( v_{b{\rm N}}-v_{\gamma {\rm N}} 
\right)\,.\label{vgeq}
\ea
In the strongly coupled regime, the velocity potentials of baryons and photons 
are almost equivalent to each other ($v_{b{\rm N}} \simeq v_{\gamma {\rm N}}$). 
After the recombination, it is a good approximation to set 
the right hand-sides of Eqs.~(\ref{vbeq})-(\ref{vgeq}) to be 0.

\section{Quasi-static approximation for sub-horizon perturbations}
\label{subsec}

In this section, we derive the second-order 
equations for the BEC and perfect-fluid density perturbations 
under a quasi-static approximation for 
the modes deep inside the Hubble radius. 
The perturbation $\delta \beta_{\rm N}$ consists of a special 
solution induced by the $\chi$-field density contrast and 
a homogenous solution which oscillates with the approximate 
frequency $2m$. 
The quasi-static approximation amounts to neglecting the 
oscillating mode of $\delta \beta_{\rm N}$ relative to 
its special solution for the perturbation dynamics 
over the cosmological time scale $H^{-1}$. 
Numerically, we will study the validity of this approximation by 
numerically solving the perturbation equations including the 
modes close to the Hubble radius.

\subsection{Analytic estimation}
\label{anasec}

The gauge-invariant density contrasts of the BEC and 
perfect fluids are defined, respectively, by 
\be
\delta_{\chi {\rm N}} \equiv \frac{\delta \rho_{\chi {\rm N}}}{\rho_\chi}\,,\qquad 
\delta_{I{\rm N}} \equiv \frac{\delta \rho_{I{\rm N}}}{\rho_I}\,.
\ee
We study the evolution of perturbations in Fourier space 
with the comoving wavenumber $k$.
{}From Eq.~(\ref{pereq1d}), we have
\be
\dot{\delta}_{\chi {\rm N}}-3\dot{\Phi}+\frac{k^2}{a^2} v_{\chi {\rm N}}
-\frac{1}{2}\dot{\delta \beta}_{\rm N}=0\,.
\label{delmeq}
\ee
Differentiating Eq.~(\ref{delmeq}) with respect to $t$ and using 
Eqs.~(\ref{pereq2d}) and (\ref{delmeq}) to eliminate 
$\dot{v}_{\chi{\rm N}}$ and $v_{\chi{\rm N}}$, it follows that 
\be
\ddot{\delta}_{\chi{\rm N}}+2H \dot{\delta}_{\chi{\rm N}}
+\frac{k^2}{a^2}\Phi-3\ddot{\Phi}-6H\dot{\Phi} 
-\frac{1}{2} \ddot{\delta \beta}_{\rm N}- H \dot{\delta \beta}_{\rm N}
-\frac{k^2}{2a^2} \delta \beta_{\rm N}=0\,.
\label{delmeq2}
\ee
In Eq.~(\ref{delbeN}), $\delta \beta_{\rm N}$ contains 
the second time derivative $\ddot{\delta}_{\chi{\rm N}}$. 
Combining Eq.~(\ref{delbeN}) with Eq.~(\ref{delmeq2}) to eliminate 
$\ddot{\delta}_{\chi{\rm N}}$, we obtain
\ba
& &
\ddot{\delta \beta}_{\rm N}+2H \dot{\delta \beta}_{\rm N}
+\left( \frac{k^2}{a^2}+4m^2 \right) \delta \beta_{\rm N}
+2 \left( \frac{k^2}{a^2}+\frac{\rho_\chi U_{,\rho \rho}}{m^2} 
\right) \delta_{\chi{\rm N}}-4H \dot{\delta}_{\chi{\rm N}} 
\nonumber \\
& &
+6 \left( \ddot{\Phi}+6H \dot{\Phi} \right)
+2\left[ 3 (3H^2+2\dot{H})-\frac{k^2}{a^2} \right] \Phi
=0\,.
\label{delbetaeq}
\ea
Taking the small-scale limit in Eq.~(\ref{delbeN}), the perturbation 
$\delta \beta_{\rm N}$ has the scale-dependence
$\delta \beta_{\rm N} \simeq -k^2/(2m^2 a^2) \delta_{\chi{\rm N}}$. 
In the linear regime of perturbation theory ($|\delta_{\chi{\rm N}}| \lesssim 0.1$), 
we require the condition $k/(ma) \lesssim 1$ to ensure that 
$|\delta \beta_{\rm N}| \lesssim 0.1$. 
In other words, the nonrelativistic BEC description 
with the background value $\beta=0$ can be approximately 
justified for the wavenumber $k/a \lesssim m$. 
Physically, this means that the BEC ground state is 
described by a coherent wave with the length scale 
$(k/a)^{-1}$ larger than the Compton wavelength $m^{-1}$ \cite{Hu:2000ke}.

In the following, we will derive the second-order differential equations 
of $\delta_{\chi {\rm N}}$ and $\delta_{I {\rm N}}$ for the 
sub-horizon modes in the range
\be
H \ll \frac{k}{a} \lesssim m\,.
\label{krange}
\ee
We impose the conditions (\ref{mcon}) to ensure the 
nonrelativistic BEC description.
For the wavenumber in the range (\ref{krange}), we exploit the 
quasi-static approximation under which the time derivatives 
$\dot{\Psi}$, $\dot{\Phi}$, and $\dot{\delta}_{\chi {\rm N}}$ are 
at most of the orders $H \Psi$, $H \Phi$, and 
$H \delta_{\chi {\rm N}}$, respectively.
In this case, Eq.~(\ref{pereq5d}) approximately reduces to
\be
\frac{k^2}{a^2}\Phi \simeq -4 \pi G \biggl( 
\rho_{\chi} \delta_{\chi {\rm N}}
-\frac{1}{2} \rho_{\chi} \delta \beta_{\rm N}
+\sum_{I} \rho_I \delta_{I {\rm N}} \biggr)\,.
\label{Phik} 
\ee
Substituting Eq.~(\ref{Phik}) into Eq.~(\ref{delbetaeq}) and using the fact 
that the term $\pi G \rho_{\chi}$ is at most of the order $H^2$, we obtain
\be
\ddot{\delta \beta}_{\rm N}+2H \dot{\delta \beta}_{\rm N}
+\left( \frac{k^2}{a^2}+4m^2 \right) \delta \beta_{\rm N} 
\simeq -2 \left( \frac{k^2}{a^2}+\frac{\rho_{\chi} U_{,\rho \rho}}{m^2} 
\right) \delta_{\chi{\rm N}}-8\pi G \sum_{I} \rho_I \delta_{I{\rm N}}\,.
\label{ddotbeta}
\ee
The general solution to Eq.~(\ref{ddotbeta}) is the sum of a special 
solution $\delta \beta_{\rm N}^{(\rs)}$ and a homogenous solution 
$\delta \beta_{\rm N}^{(\rH)}$, i.e., 
\be
\delta \beta_{\rm N}=\delta \beta_{\rm N}^{(\rs)}+
\delta \beta_{\rm N}^{(\rH)}\,.
\label{delbetaso}
\ee
The special solution can be derived by neglecting the time 
derivatives on the left hand-side of Eq.~(\ref{ddotbeta}), 
such that 
\be
\delta \beta_{\rm N}^{(\rs)}=
-\frac{2(k^2+a^2 \rho_{\chi} U_{,\rho \rho}/m^2)\delta_{\chi {\rm N}}
+8 \pi G a^2 \sum_I \rho_I \delta_{I{\rm N}}}{k^2+4m^2 a^2}\,.
\label{delbs}
\ee
Since $\delta \beta_{\rm N}^{(\rs)}$ is directly related to 
the matter density contrasts, the typical time scale for 
its variation should be of order the Hubble time $H^{-1}$. 

The homogenous solution can be obtained by setting the terms 
on the right hand-side of Eq.~(\ref{ddotbeta}) to be 0. 
Under the condition (\ref{krange}), the dominant contribution 
to the frequency $\omega_k=\sqrt{k^2/a^2+4m^2}$ in 
Eq.~(\ref{ddotbeta}) is the mass term $2m$. 
This means that $\delta \beta_{\rm N}^{(\rH)}$ oscillates with 
the approximate time period $(2m)^{-1}$, which is much shorter 
than $H^{-1}$ for $m \gg H$.
The WKB solution to this oscillating mode is given by 
\be
\delta \beta_{\rm N}^{(\rH)}=\frac{1}{a\sqrt{2\omega_k}} 
\left[ A\cos \left( \int_0^t \omega_k {\rm d}\tilde{t} \right)
+B \sin \left( \int_0^t \omega_k {\rm d}\tilde{t} \right) \right]\,,
\label{delbeos}
\ee
where $A$ and $B$ are integration constants. 
Under the initial condition $\delta \beta_{\rm N}=\delta \beta_{\rm N}^{(\rs)}+
\delta \beta_{\rm N}^{(\rH)}=0$ at $t=0$, 
the constant $A$ is fixed to be 
$A=-a_i \sqrt{2\omega_{ki}}\,\delta \beta_{{\rm N}i}^{(\rs)}$, 
where $a_i$, $\omega_{ki}$, and $\delta \beta_{{\rm N}i}^{(\rs)}$
are the initial values of $a$, $\omega_k$, and 
$\delta \beta_{{\rm N}}^{(\rs)}$ respectively. 
Imposing the initial condition $\dot{\delta \beta}_{\rm N}=0$ 
at $t=0$ further, 
we obtain the constant $B$ which is suppressed by the 
factor $H/m$ in comparison to $A$. 
Neglecting the term $B\sin ( \int_0^t \omega_k {\rm d}\tilde{t})$ 
in Eq.~(\ref{delbeos}), the homogenous solution yields
\be
\delta \beta_{\rm N}^{(\rH)} \simeq 
-\frac{a_i}{a} \sqrt{\frac{\omega_{ki}}{\omega_k}}
\delta \beta_{{\rm N}i}^{(\rs)} \cos 
\left( \int_0^t \omega_k {\rm d}\tilde{t} \right)\,.
\label{delHso}
\ee
Thus, the contributions $\delta \beta_{\rm N}^{(\rs)}$ 
and $\delta \beta_{\rm N}^{(\rH)}$ to the total solution 
(\ref{delbetaso}) are given, respectively, by 
Eqs.~(\ref{delbs}) and (\ref{delHso}).
In the regime $k/a \ll m$, the solution reduces to 
$\delta \beta_{\rm N} \simeq \delta
\beta_{\rm N}^{(\rs)}-(a_i/a)\delta \beta_{{\rm N}i}^{(\rs)}
\cos(2mt)$. While $\beta=0$ at the background level, 
the perturbation $\delta \beta_{\rm N}$ contains the matter-induced 
mode $\delta\beta_{\rm N}^{(\rs)}$ slowly varying over the Hubble
time scale $H^{-1}$ as well as the oscillating mode 
$\delta \beta_{\rm N}^{(\rH)}$ rapidly changing over 
the time scale $(2m)^{-1}$.

Applying the quasi-static approximation to Eq.~(\ref{delmeq2}) 
for the gravitational potential $\Phi$ 
deep inside the Hubble radius and using Eq.~(\ref{Phik}), 
it follows that 
\be
\ddot{\delta}_{\chi{\rm N}}+2H \dot{\delta}_{\chi{\rm N}}
-4 \pi G \left( \rho_\chi \delta_{\chi{\rm N}}
+\sum_{I} \rho_I \delta_{I {\rm N}} \right)
-\frac{1}{2} \ddot{\delta \beta}_{\rm N}- H \dot{\delta \beta}_{\rm N}
-\frac{k^2}{2a^2} \delta \beta_{\rm N} \simeq 0\,,
\label{delm2}
\ee
where we ignored the second term on the right hand-side of 
Eq.~(\ref{Phik}) relative to $k^2/(2a^2)\delta \beta_{\rm N}$.
The time derivatives $\ddot{\delta \beta}_{\rm N}^{(\rs)}$ and 
$H \dot{\delta \beta}_{\rm N}^{(\rs)}$ present in Eq.~(\ref{delm2}) 
are at most of order $H^2 \delta \beta_{\rm N}^{(\rs)}$, so 
we can neglect these contributions 
relative to the term $(k^2/2a^2)\delta \beta_{\rm N}^{(\rs)}$. 
For the homogenous solution $\delta \beta_{\rm N}^{(\rH)}$, 
the term $H \dot{\delta \beta}_{\rm N}^{(\rH)}$ is suppressed relative to 
$\ddot{\delta \beta}_{\rm N}^{(\rH)}/2$.
On using the property $\ddot{\delta \beta}_{\rm N}^{(\rH)} \simeq 
-\omega_k^2  \delta \beta_{\rm N}^{(\rH)}$, we have
$-\ddot{\delta \beta}_{\rm N}^{(\rH)}/2
-k^2/(2a^2) \delta \beta_{\rm N}^{(\rH)} \simeq 2m^2 
\delta \beta_{\rm N}^{(\rH)}$. 
Then, Eq.~(\ref{delm2}) reduces to 
\be
\ddot{\delta}_{\chi{\rm N}}+2H \dot{\delta}_{\chi{\rm N}}
-4 \pi G \biggl( \rho_\chi \delta_{\chi{\rm N}}
+\sum_{I} \rho_I \delta_{I {\rm N}} \biggr)
-\frac{k^2}{2a^2} \delta \beta_{\rm N}^{(\rs)}
-2m^2 \frac{a_i}{a} \sqrt{\frac{\omega_{ki}}{\omega_k}}
\delta \beta_{{\rm N}i}^{(\rs)} \cos 
\left( \int_0^t \omega_k {\rm d}\tilde{t} \right)
\simeq 0\,.
\label{delmos}
\ee
Initially, the last term on the left hand-side 
of Eq.~(\ref{delmos}) is larger than $-k^2/(2a^2)\delta \beta_{\rm N}^{(\rs)}$, 
but the former oscillates between $+1$ and $-1$ with the approximate 
frequency $2m$. Taking the time average over the Hubble time 
scale in Eq.~(\ref{delmos}) and using the special solution (\ref{delbs}), we obtain
\be
\ddot{\delta}_{\chi{\rm N}}+2H  \dot{\delta}_{\chi{\rm N}}
+\left( c_s^2 \frac{k^2}{a^2}-4\pi G \rho_{\chi} \right) \delta_{\chi {\rm N}}
-4\pi G \left( 1+\frac{k^2}{4m^2 a^2} \right)^{-1} 
\sum_I \rho_I \delta_{I{\rm N}} \simeq 0\,,
\label{delm3}
\ee
where $c_s^2$ is the effective sound speed squared defined by 
\be
c_s^2 \equiv \frac{k^2}{4m^2 a^2+k^2}
+\frac{a^2 \rho_{\chi} U_{,\rho \rho}}
{m^2(4m^2 a^2+k^2)}\,.
\label{cs}
\ee

The formation of a coherent BEC state gives rise to a quantum pressure
with the scale-dependent sound speed squared 
$c_{s1}^2=k^2/(4m^2 a^2+k^2)$. 
This value of $c_{s1}^2$ coincides with the one derived in Ref.~\cite{Hwang:2009js} 
by considering the perturbation $\delta \phi$ of a massive axion field $\phi$ 
and taking the time average over the background axion oscillations. 
This shows that our nonrelativistic BEC description based on the 
Madelung representation (\ref{chiMa}) is consistent with 
the approach of scalar-field perturbations in the rapidly 
oscillating regime with $m \gg H$. 
In the limit that $k/a \gg m$, $c_{s1}^2$ approaches the value 1, 
but, for scales much larger than the Compton wavelength 
($k/a \ll m$), it follows that $c_{s1}^2 \simeq k^2/(4m^2 a^2) \ll 1$. 
The fact that this latter sound speed gives rise to a quantum pressure 
due to the uncertainty principle was originally recognized 
in Ref.~\cite{Lif}.
With the BEC formation, there is a critical Jeans-scale wavenumber 
$k_J$ at which the  gravitational interaction $4\pi G \rho_{\chi}$ 
balances the pressure term $c_{s1}^2 k^2/a^2$. 
In the regime $k/a \ll m$, we have
\be
k_J= a\left( 16\pi G m^2 \rho_{\chi} \right)^{1/4}\,.
\label{kJ}
\ee
For $k>k_J$, the quantum pressure suppresses the gravitational 
instability of $\delta_{\chi {\rm N}}$.

The self-coupling potential $U(\rho)$ with a repulsive ($U_{,\rho \rho}>0$) 
or attractive ($U_{,\rho \rho}<0$) interaction leads to the suppressed 
or enhanced growth of $\delta_{\chi{\rm N}}$
through the second sound speed squared 
$c_{s2}^2=a^2 \rho_{\chi} U_{,\rho \rho}/[m^2(4m^2 a^2+k^2)]$ 
in Eq.~(\ref{cs}). For the self-coupling potential 
$U(\rho)=\lambda \rho^2/4$ of the two-body interaction, 
we have $c_{s2}^2=\lambda \rho_{\chi}/(8m^4)$ for
$k/a \ll m$. This agrees with the expression derived in 
Refs.~\cite{Chavanis:2011uv,Suarez:2016eez,Desjacques:2017fmf}.
Our result of $c_{s2}^2$ can be applied to the general 
self-interacting potential $U(\rho)$ as well as to 
the wavenumber $k/a$ close to $m$. 
In the regime $k/a \ll m$, the critical wavenumber $k_S$ at which 
$4\pi G \rho_{\chi}$ balances $|c_{s2}^2| k^2/a^2$ is given by 
\be
k_S=a \left( \frac{16 \pi G m^4}{|U_{,\rho \rho}|} 
\right)^{1/2}\,.
\label{kS}
\ee
For the modes $k>k_S$, the gravitational growth of $\delta_{\chi{\rm N}}$ 
is modified by the self-coupling potential.

There is also the critical wavenumber $k_I$ at which $c_{s1}^2$ and 
$c_{s2}^2$ on the right hand-side of Eq.~(\ref{cs}) have 
the same amplitudes, i.e., 
\be
k_I=a \left( \frac{\rho_{\chi} |U_{,\rho \rho}|}{m^2} 
\right)^{1/2}\,.
\label{kI}
\ee
For $k>k_I$, the quantum pressure dominates over the self-interaction. 
In Sec.~\ref{selfsec}, we will study the evolution of $\delta_{\chi{\rm N}}$ 
by comparing the three critical wavenumbers $k_J$, $k_S$, and $k_I$.

In the regime $k/a \ll m$, the last term on the left hand-side of 
Eq.~(\ref{delm3}) reduces to the standard form 
$-4\pi G \sum_I \rho_I \delta_{I{\rm N}}$. 
In the presence of CDM and baryons, their density contrasts 
$\delta_{c{\rm N}}$ and $\delta_{b{\rm N}}$ affect the evolution 
of $\delta_{\chi{\rm N}}$ through the gravitational interaction 
mediated by $\Phi$.
Let us derive the second-order differential equations of 
$\delta_{c{\rm N}}$ and $\delta_{b{\rm N}}$ 
under the same approximation scheme as $\delta_{\chi{\rm N}}$.
Differentiating Eq.~(\ref{pereq3d}) with respect to $t$ and using 
Eq.~(\ref{pereq4d}), we find
\ba
& &
\ddot{\delta}_{I{\rm N}}+\left( 2+3c_I^2-6w_I \right) H\dot{\delta}_{I{\rm N}}
+\left[ c_I^2 \frac{k^2}{a^2}+3 \left( 5H^2+\dot{H} \right) \left( c_I^2-w_I \right) 
+3H (c_I^2)^{\cdot} \right] \delta_{I{\rm N}} \nonumber \\
& &
+(1+w_I) \left[ \frac{k^2}{a^2} \Phi+3H \left( 3c_I^2-2 \right) 
\dot{\Phi}-3\ddot{\Phi} \right]=0\,.
\label{dceq}
\ea
For the modes deep inside the Hubble radius, the terms
$3H (3c_I^2-2)\dot{\Phi}-3\ddot{\Phi}$ in Eq.~(\ref{dceq}) 
are negligible relative to $(k^2/a^2)\Phi$. 
This latter Laplacian term, which is approximately given by Eq.~(\ref{Phik}), 
contains the contribution $-4\pi G \sum \rho_I \delta_{I{\rm N}}$ 
responsible for the gravitational clustering of $\delta_{I{\rm N}}$.
In Eq.~(\ref{dceq}), there exists the term 
$c_I^2 (k^2/a^2) \delta_{I{\rm N}}$ preventing the 
growth of $\delta_{I{\rm N}}$ on scales smaller the 
sound horizon ($c_I k/a>H$). 
For $c_I^2={\cal O}(1)$, which is typically the case for DE 
perturbations, the density contrast 
does not grow for most of the modes inside the Hubble radius. 
On the other hand, the sound speeds of CDM and baryons 
are much smaller than 1 after the recombination epoch, so 
$\delta_{c{\rm N}}$ and $\delta_{b{\rm N}}$ are subject to 
the gravitational instabilities.
In the following, we take the nonrelativistic limits,
\be
c_I^2 \to 0\,,\quad w_I \to 0 \qquad 
{\rm for} \quad I=c,b\,.
\label{cIwI}
\ee
We substitute Eq.~(\ref{Phik}) into Eq.~(\ref{dceq}) with the 
approximation $\delta \beta_{\rm N} \simeq \delta \beta_{\rm N}^{(\rs)}$. 
The oscillating mode $\delta \beta_{\rm N}^{(\rH)}$ in Eq.~(\ref{delHso})
should not contribute to the growth of $\delta_{I{\rm N}}$ over the 
Hubble time scale. 
Then, the CDM and baryon density contrasts for the modes 
deep inside the Hubble radius obey 
\be
\ddot{\delta}_{I{\rm N}}+2H\dot{\delta}_{I{\rm N}}
-4\pi G \sum_{I} \rho_I \delta_{I{\rm N}}
-4\pi G \rho_{\chi} \left( 1+c_s^2 \right) \delta_{\chi {\rm N}} 
\simeq 0\,,
\qquad 
{\rm for} \quad I=c,b\,,
\label{delNeq}
\ee
where $c_s^2$ is given by Eq.~(\ref{cs}). 
To our knowledge, the contribution of $c_s^2$ to $\delta_{I{\rm N}}$ 
in the form (\ref{delNeq}) was not recognized in the literature. 
For the wavenumber $k/a \ll m$, we have 
$c_s^2 \simeq k^2/(4m^2 a^2) \ll 1$ and hence 
the last term on the left hand-side of Eq.~(\ref{delNeq}) 
approximately reduces to $-4\pi G \rho_{\chi} \delta_{\chi {\rm N}}$.
The evolution of $\delta_{\chi {\rm N}}$ is affected by those of 
$\delta_{c{\rm N}}$ and $\delta_{b{\rm N}}$ through 
Eqs.~(\ref{delm3}) and (\ref{delNeq}), and vice versa. 

\subsection{Numerical solutions}
\label{nusec}

In order to confirm the accuracy of approximations 
exploited in Sec.~\ref{anasec}, we numerically solve the perturbation 
equations of motion during the matter era in which 
the BEC energy density dominates over the other matter densities.
In this section we focus on the case without BEC 
self-interactions ($U=0$), but in Sec.~\ref{selfsec} 
we will take into account the self-coupling potential as well as 
the perturbations of CDM and baryons. 
We integrate the perturbation Eqs.~(\ref{pereq2d}), (\ref{pereq5d})-(\ref{pereq6d}), 
(\ref{delmeq}), (\ref{delbetaeq}) with (\ref{pereq8d}), along with 
the background Eqs.~(\ref{coneq2}) and 
(\ref{back1})-(\ref{back2}).

\begin{figure}[h]
\begin{center}
\includegraphics[height=3.1in,width=3.5in]{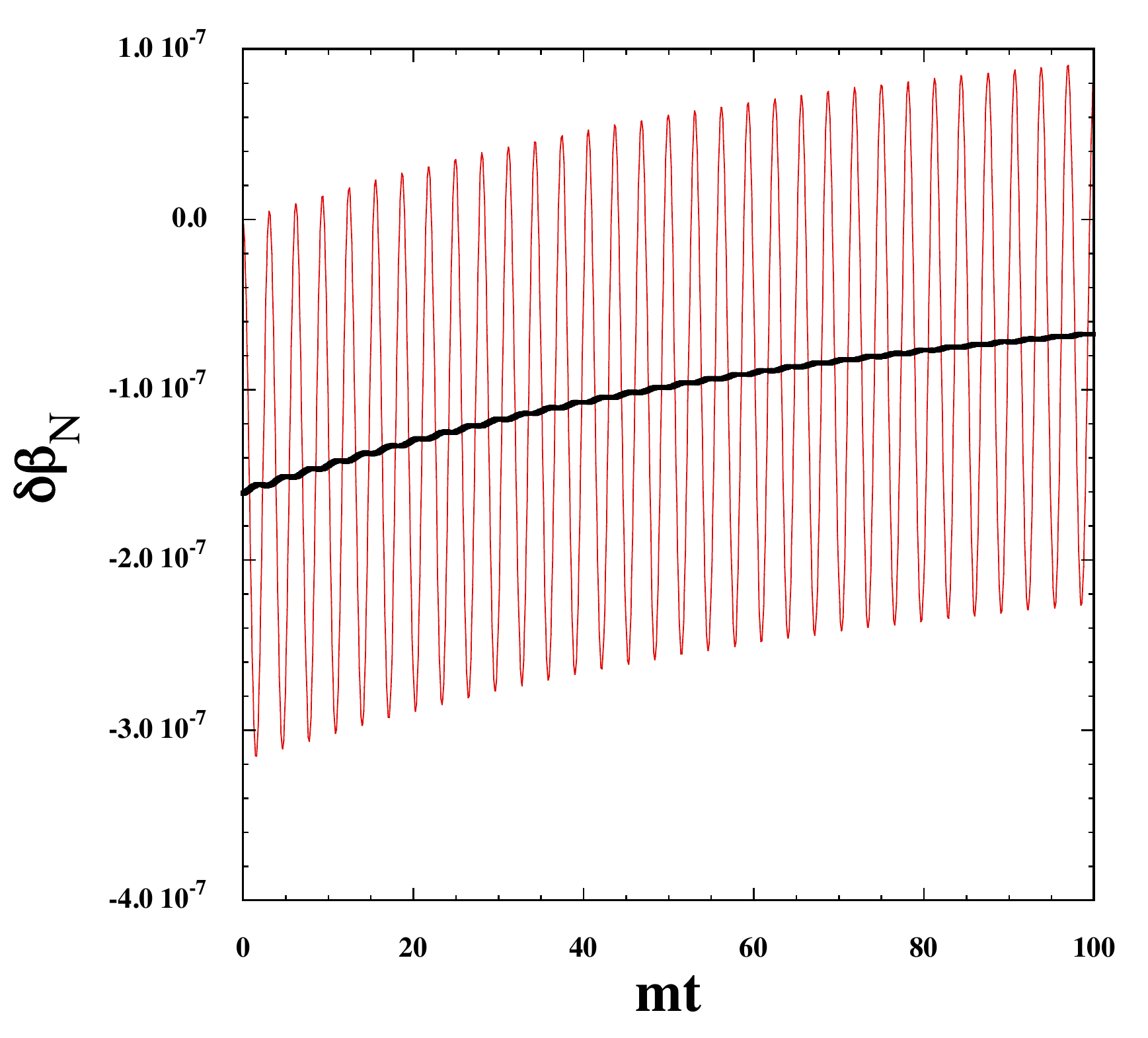}
\includegraphics[height=3.1in,width=3.5in]{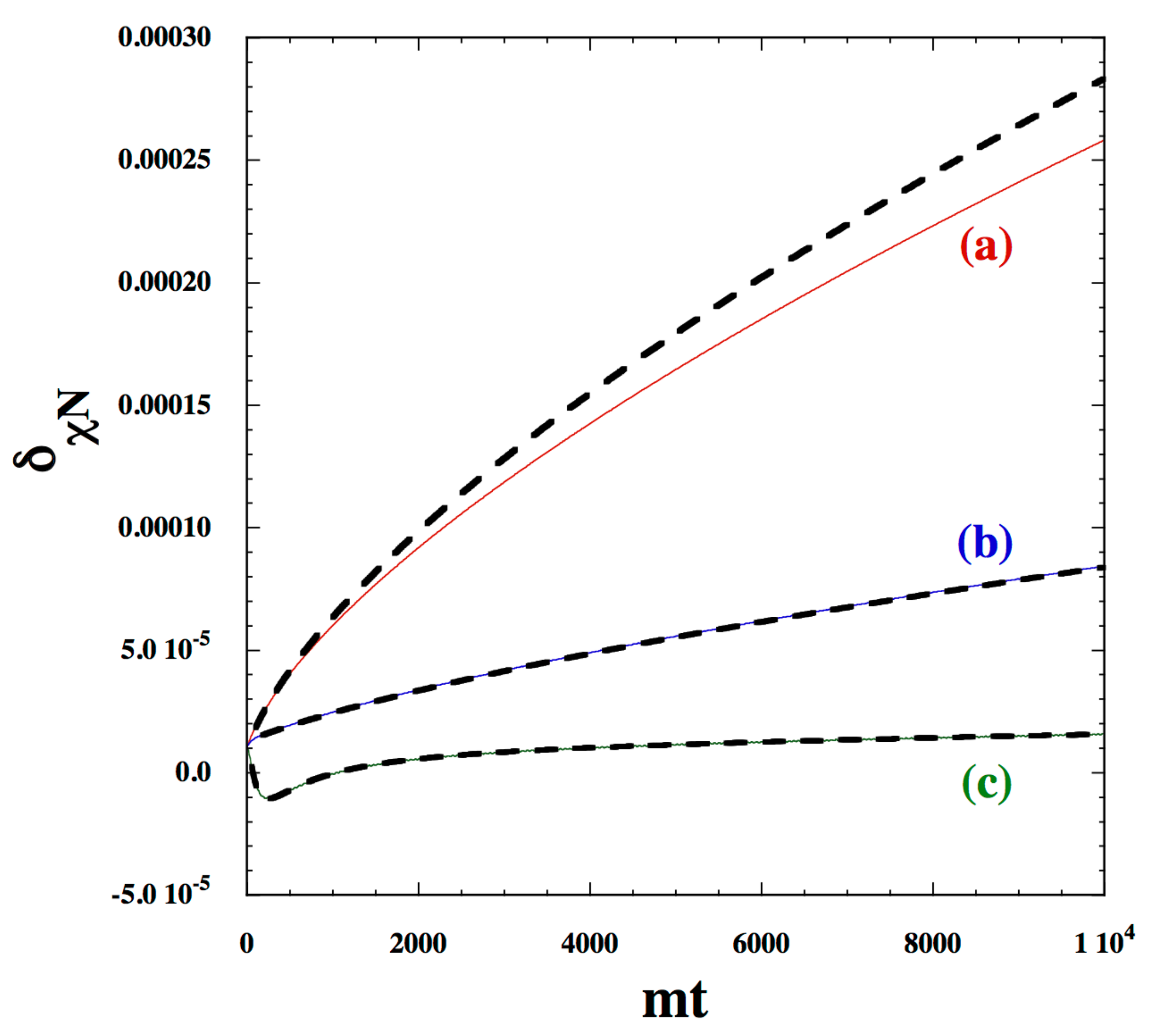}
\end{center}
\caption{\label{fig1}
(Left) The red line shows the evolution of $\delta \beta_{\rm N}$
versus $mt$ during the matter era dominated by 
the BEC energy density, without the self-coupling potential $U$.
The initial conditions are chosen to be $H=10^{-2}m$, 
$\rho_{\chi}=3H^2/(8\pi G)$, 
$\delta_{\chi{\rm N}}=\dot{\delta}_{\chi{\rm N}}/H=10^{-5}$, 
$\delta \beta_{{\rm N}}=0$, $\dot{\delta \beta}_{\rm N}=0$, 
and ${\cal K}=k/(aH)=18$ at $t=0$. The thick black line 
is the evolution of the special solution 
$\delta \beta_{\rm N}^{(\rs)}$ given by Eq.~(\ref{delbs}).
(Right) The red, blue, and green lines correspond to the evolutions 
of $\delta_{\chi{\rm N}}$ for three different wavenumbers: (a) ${\cal K}=3$, 
(b) ${\cal K}=18$, and (c) ${\cal K}=30$ at $mt=0$, respectively. 
The other initial conditions are the same as those used in the 
left panel. The thick dashed plots are derived by solving 
the approximate Eq.~(\ref{delm3}). 
This approximation is accurate for the perturbations 
deep inside the Hubble radius (${\cal K} \gg 1$).} 
\end{figure}

In the left panel of Fig.~\ref{fig1}, we plot the evolution of 
$\delta \beta_{\rm N}$ for the initial conditions $H=10^{-2}m$ and 
$\delta \beta_{{\rm N}}=\dot{\delta \beta}_{\rm N}=0$. 
Since $H$ decreases in time, the nonrelativistic 
condition $H \ll m$ is always satisfied. 
We choose the wavenumber $k$ to be ${\cal K} \equiv k/(aH)=18$ at $t=0$, 
in which case the perturbation is deep inside the Hubble radius 
during the matter era. 
Since this mode is in the regime $k/a \ll m$, the special solution 
(\ref{delbs}) and the homogenous solution (\ref{delHso}) reduce, 
respectively, to 
\be
\delta \beta_{{\rm N}}^{(\rs)} \simeq -\frac{k^2}{2m^2 a^2} 
\delta_{\chi {\rm N}}\,,\qquad
\delta \beta_{{\rm N}}^{(\rH)} \simeq -\frac{a_i}{a}
\delta \beta_{{\rm N}i}^{(\rs)} \cos(2mt)\,,
\ee
where $\delta \beta_{\rm N}$ is the sum of these two modes.
In Fig.~\ref{fig1}, we find that $\delta \beta_{{\rm N}}$ 
oscillates with the period $\pi/m$ around the slowly 
varying central value $\delta \beta_{{\rm N}}^{(\rs)}$ due to 
the homogenous mode $\delta \beta_{{\rm N}}^{(\rH)}$, with 
an amplitude related to the initial value $\delta \beta_{{\rm N}i}^{(\rs)}$. 
The special solution $\delta \beta_{{\rm N}}^{(\rs)}$ shown as 
a thick black line in Fig.~\ref{fig1}, which is proportional to $\delta_{\chi {\rm N}}$, 
exhibits only a tiny oscillation with an amplitude much smaller than 
the time-averaged value $-k^2/(2m^2 a^2)\delta_{\chi {\rm N}}$. 
It should be a good approximation to neglect the homogenous oscillating 
mode $\delta \beta_{{\rm N}}^{(\rH)}$ relative to 
$\delta \beta_{{\rm N}}^{(\rs)} \simeq -k^2/(2m^2 a^2)\delta_{\chi {\rm N}}$ 
for the evolution of $\delta_{\chi {\rm N}}$ on cosmological time scales.

In the right panel of Fig.~\ref{fig1}, we show the evolutions of 
$\delta_{\chi{\rm N}}$ for three different wavenumbers with 
the same initial conditions as those used in the left. 
The case (a) corresponds to the wavenumber ${\cal K}=k/(aH)=3$ 
at $t=0$. The density contrast $\delta_{\chi {\rm N}}$ grows 
by the gravitational source term $-4\pi G \rho_{\chi} \delta_{\chi{\rm N}}$ 
in Eq.~(\ref{delm3}). For this mode 
the Laplacian term $c_s^2 (k^2/a^2) \delta_{\chi{\rm N}}$ is smaller than  
$4\pi G \rho_{\chi} \delta_{\chi {\rm N}}$, so the growth of 
$\delta_{\chi{\rm N}}$ is hardly prevented by the quantum pressure. 
The thick dashed line above the red line (a) in Fig.~\ref{fig1} 
is obtained by integrating the approximate Eq.~(\ref{delm3}) 
with Eq.~(\ref{cs}). In case (a), the approximate solution exhibits some 
difference from the numerical solution. This property is mostly 
attributed to the fact that, for the modes close to the Hubble radius, 
the terms of order $H^2 \Phi$ cannot be ignored 
relative to $(k^2/a^2)\Phi$ in Eq.~(\ref{pereq5d}).
Indeed, implementing such contributions to Eq.~(\ref{pereq5d}) gives rise 
to the gravitational potential $\Phi$ smaller than that estimated by 
Eq.~(\ref{Phik}). This is the main reason why the full numerical solution (a) 
of $\delta_{\chi{\rm N}}$ is smaller than the
analytic estimation (\ref{delm3}) derived for the modes ${\cal K} \gg 1$.

The case (b) in Fig.~\ref{fig1} corresponds to the wavenumber 
${\cal K}=18$ at $t=0$. We observe that the growth of 
$\delta_{\chi {\rm N}}$ is suppressed in comparison to case (a). 
On using the Friedmann equation $3H^2 \simeq 8 \pi G \rho_{\chi}$ 
in Eq.~(\ref{kJ}), the critical wavenumber associated with 
the Jeans scale can be estimated as 
\be
{\cal K}_J \equiv \frac{k_J}{aH} \simeq 1.6 \sqrt{\frac{m}{H}}\,.
\ee
With the initial Hubble parameter $H=10^{-2}m$, we have 
${\cal K}_J \simeq 16$ and hence the mode ${\cal K}=18$ 
is affected by the quantum pressure. 
The suppressed growth of $\delta_{\chi{\rm N}}$ starts to 
be at work for the initial value of ${\cal K}$ close to ${\cal K}_J$.
Especially for ${\cal K} \gg {\cal K}_J$, the quantum pressure 
leads to the strong suppression of $\delta_{\chi{\rm N}}$.
This property can be confirmed in case (c) of Fig.~\ref{fig1}, 
which corresponds to the initial wavenumber ${\cal K}=30$.

In cases (b) and (c) the evolutions of 
$\delta_{\chi{\rm N}}$ obtained by solving the approximate Eq.~(\ref{delm3})
show good agreement with the full numerical results, by reflecting the 
fact that the perturbations are always in the regime ${\cal K} \gg 1$.
Thus, the approximate second-order differential Eq.~(\ref{delm3}) 
can be trustable for the modes deep inside the Hubble radius. 
In the right panel of Fig.~\ref{fig1}, we also observe that the oscillating mode 
in $\delta \beta_{\rm N}$ does not give rise to any large oscillations of 
$\delta_{\chi{\rm N}}$.

\section{BEC self-interactions}
\label{selfsec}

In this section, we study the effect of BEC self-interactions 
on the dynamics of cosmological perturbations. 
For concreteness, we consider the self-coupling potential 
\be
U(\rho)=\frac{1}{4} \lambda \rho^2\,,
\label{Urho}
\ee
where $\lambda$ is a coupling constant. 
In Eq.~(\ref{betaeq}), the contribution to $\beta$ arising from 
$U(\rho)$ is given by 
\be
\beta_U \equiv -\frac{U_{,\rho}}{m^2}
=-\frac{\lambda \rho_{\chi}}{4m^4}
=-\frac{3}{2} \beta_m \mu \Omega_{\chi}\,,
\label{betaU}
\ee
where 
\be
\beta_m \equiv \frac{H^2}{m^2}\,,\qquad
\mu \equiv \frac{\lambda M_{\rm pl}^2}{2m^2}\,,\qquad
\Omega_{\chi} \equiv \frac{\rho_{\chi}}{3M_{\rm pl}^2 H^2}\,.
\label{mudef}
\ee
We require that both $|\beta_U|$ and $\beta_m$ are smaller than 
the order 1 to ensure the nonrelativistic BEC description. 
The ratio between $U$ and $\rho_{\chi}$ is given by 
\be
\frac{U}{\rho_{\chi}}=
-\frac{1}{4} \beta_{U}\,.
\ee
Under the condition $|\beta_{U}| \ll 1$, 
the self-coupling potential is suppressed 
in comparison to $\rho_{\chi}=2m^2 \chi^* \chi$.

For a real scalar field $\phi=\sqrt{2} \chi$ with the 
axion-type potential
\be
V=m^2 f^2 \left[ 1-\cos \left( \frac{\phi}{f} 
\right) \right]\,,
\label{Vpo}
\ee
the expansion of $V$ around $\phi=0$ leads to 
\be
V=m^2 \chi^2-\frac{m^2}{6f^2}\chi^4+{\cal O} (\chi^6)\,.
\label{Vpo2}
\ee
The first contribution to $V$ in Eq.~(\ref{Vpo2}) is the mass Lagrangian 
$m^2 \chi^* \chi$ in Eq.~(\ref{SMtotal}), while the second one 
corresponds to the self-coupling potential (\ref{Urho}) with 
$\lambda=-2m^2/(3f^2)$. 
In this case, the dimensionless constant $\mu$ defined in 
Eq.~(\ref{mudef}) reads
\be
\mu=-\frac{M_{\rm pl}^2}{3f^2}\,.
\label{mure}
\ee
Since $\mu$ is negative, the axion potential of the form 
(\ref{Vpo}) leads to an attractive 
self-interaction \cite{Guth:2014hsa,Desjacques:2017fmf}. 
For $f \ll M_{\rm pl}$, $|\mu|$ is larger than the order 1. 

We study the evolution of density contrasts 
for the total  potential of the form 
$V=m^2 \chi^* \chi+\lambda (\chi^* \chi)^2/4$, 
but our analysis also covers the axion-type potential 
(\ref{Vpo}) expanded around its minimum. 
We will focus on the case $\mu<0$ 
in the following discussion.

\subsection{Scales relevant to the self-coupling}

At the onset of BEC formation, which is expressed by 
the subscript $*$ for time-dependent quantities, 
the variable (\ref{betaU}) reads
\be
\beta_{U*}=-\frac{3}{2} \beta_{m*} \mu
\Omega_{\chi *}\,,
\ee
where 
\be
\beta_{m*}=\frac{H_0^2}{m^2}
\frac{\Omega_{M0} (a_*+a_{\rm eq})}{a_*^4}\,,\qquad
\Omega_{\chi *}= \frac{\rho_{\chi *}}
{3M_{\rm pl}^2 H_*^2}
=\frac{\Omega_{\chi 0}}{\Omega_{M 0}}
\frac{a_*}{a_*+a_{\rm eq}}\,.
\label{Omechi}
\ee
In Eq.~(\ref{Omechi}), we used Eq.~(\ref{Ha}) and neglected 
the contribution of DE to the Hubble expansion rate.
Since $|\beta_{U}|$ decreases as $|\beta_{U}| \propto a^{-3}$ 
for $a>a_*$, the nonrelativistic BEC description can be 
ensured for $|\beta_{U*}| \ll 1$. 
This gives an upper limit on the self-coupling strength, as
\be
|\mu| \ll \frac{2}{3} \frac{1}{ \beta_{m*}} \frac{\Omega_{M0}}
{\Omega_{\chi 0}} \frac{a_*+a_{\rm eq}}{a_*}\,.
\label{mucon1}
\ee

\begin{figure}[h]
\begin{center}
\includegraphics[height=3.0in,width=4.5in]{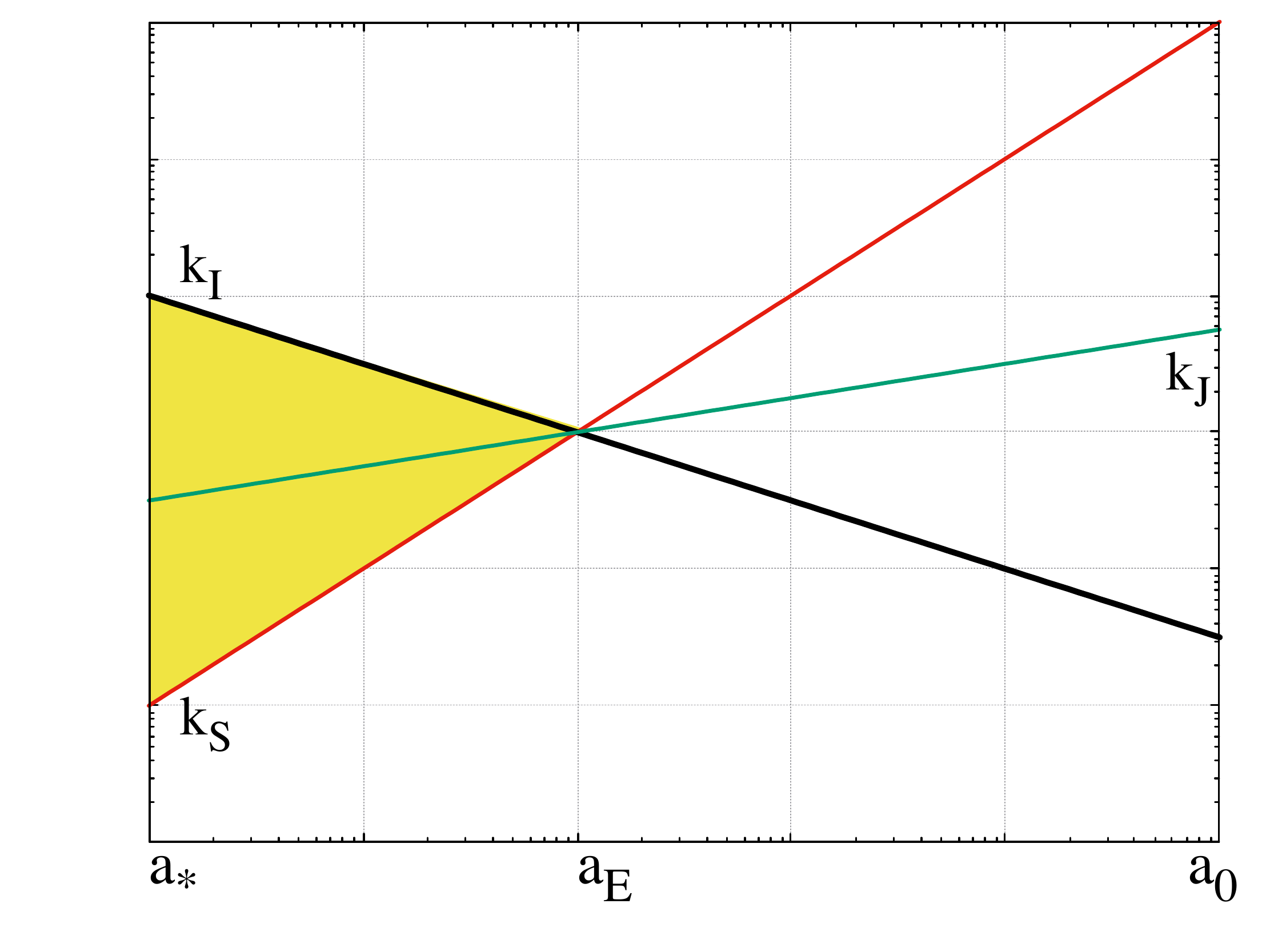}
\end{center}
\caption{
Evolution of $k_J$, $k_S$, and $k_I$ versus $a$ with the logarithmic scales 
for both vertical and horizontal axes. 
The scale factors $a_*$, $a_{\rm E}$, and $a_0~(=1)$ are those 
at the onset of BEC, $k_J=k_S=k_I$, and today, respectively. 
In the yellow shaded region, the effect of self-coupling 
on the frequency $\omega_\chi^2=c_s^2 k^2/a^2-4\pi G \rho_{\chi}$ 
of $\delta_{\chi {\rm N}}$ dominates over both the quantum pressure 
and the gravitational instability term $-4\pi G \rho_{\chi}$, 
see Eq.~(\ref{omese}).
\label{fig2}} 
\end{figure}

For the modes deep inside the Hubble radius, the density contrast 
$\delta_{\chi {\rm N}}$ approximately obeys Eq.~(\ref{delm3}).
For the self-coupling potential (\ref{Urho}), the sound speed 
squared (\ref{cs}) is given by  
\be
c_s^2= \left( \frac{k^2}{4m^2 a^2}-\frac{1}{2} \beta_U \right) 
\left( 1+\frac{k^2}{4m^2 a^2} \right)^{-1}\,.
\label{csbeta}
\ee
For the wavenumber in the range $k/a \ll m$, 
Eq.~(\ref{csbeta}) reduces to $c_s^2 \simeq k^2/(4m^2 a^2)-\beta_U/2$.
For $k<k_I$, where $k_I$ is given by Eq.~(\ref{kI}), the self-coupling 
term $-\beta_U/2$ is the dominant contribution to $c_s^2$. 
Since we are considering the case $\mu<0$, 
the term $-\beta_U/2$ is negative. 
This can induce the Laplacian instability of $\delta_{\chi {\rm N}}$ 
for some particular values of $k$. 
As we estimated in Sec.~\ref{anasec}, whether the Laplacian instability 
is present or not is the comparison of the $c_s^2 k^2/a^2$ term 
with those responsible for the gravitational instability. 
On using the property $\rho_{\chi}=\rho_{\chi 0}a^{-3}$, where 
$\rho_{\chi 0}=3M_{\rm pl}^2 H_0^2 \Omega_{\chi 0}$ is today's 
energy density of the BEC, the three critical wavenumbers 
(\ref{kJ}), (\ref{kS}), and (\ref{kI}) can be expressed, respectively, as
\ba
k_J &=& 5.22 \times 10^{-4}\,h\,\Omega_{\chi 0}^{1/4}
\left( \frac{m}{H_0} \right)^{1/2} a^{1/4}~~{\rm Mpc}^{-1}\,,
\label{kJ2}\\
k_S &=& 4.72 \times 10^{-4}\,h |\mu|^{-1/2} 
\frac{m}{H_0}a~~{\rm Mpc}^{-1}\,.\\
k_I &=& 5.78 \times 10^{-4}\,h\,\Omega_{\chi 0}^{1/2} 
|\mu|^{1/2} a^{-1/2}~~{\rm Mpc}^{-1}\,.
\label{kI2}
\ea
In the asymptotic past ($a \to 0$), the largest wavenumber 
is $k_I$, while the smallest one is $k_S$. 
Let us consider the case in which the inequality 
$k_I>k_J>k_S$ is satisfied at $a=a_*$.
As we see in Fig.~\ref{fig2}, there is a moment at which 
$k_I$, $k_J$, and $k_S$ become equivalent to each other. 
This corresponds to the instant
$\mu^2 \rho_{\chi}=2m^2 M_{\rm pl}^2$, which translates 
to the scale factor   
\be
a_{\rm E}=\left( \frac{3\mu^2 H_0^2 \Omega_{\chi 0}}{2m^2} 
\right)^{1/3}\,,
\label{aS2}
\ee
with the wavenumber
\be
k_{\rm E}=5.40 \times 10^{-4}\,h\,\Omega_{\chi 0}^{1/3} 
|\mu|^{1/6} \left( \frac{m}{H_0} \right)^{1/3}~~{\rm Mpc}^{-1}\,.
\label{kE}
\ee
The necessary condition for the Laplacian instability to 
occur is given by 
\be
a_{\rm E}>a_*\quad \to \quad |\mu|>
\frac{m}{H_0} \sqrt{\frac{2a_*^3}{3\Omega_{\chi 0}}}\,.
\label{mucon2}
\ee
Under this condition, for the modes in the range $k_S<k<k_I$, 
the negative Laplacian term $c_s^2 k^2/a^2$ associated with 
the self-coupling dominates 
over the term $-4\pi G \rho_{\chi}$  
during the time interval $a_*<a<a_{\rm E}$. 
This is shown as a yellow shaded region in Fig.~\ref{fig2}.
On using Eq.~(\ref{Omechi}) and combining the bound (\ref{mucon2}) 
with (\ref{mucon1}), it follows that 
\be
\sqrt{\frac{2}{3} \frac{1}{ \beta_{m*}} \frac{\Omega_{M0}}
{\Omega_{\chi 0}} \frac{a_*+a_{\rm eq}}{a_*}}
<|\mu| \ll \frac{2}{3} \frac{1}{ \beta_{m*}} \frac{\Omega_{M0}}
{\Omega_{\chi 0}} \frac{a_*+a_{\rm eq}}{a_*}\,.
\label{Omein}
\ee
For the existence of $|\mu|$ in this range, 
we require the condition 
\be
\frac{\Omega_{\chi 0}}{\Omega_{M0}}
\ll \frac{2}{3}\frac{1}{\beta_{m*}} \frac{a_*+a_{\rm eq}}{a_*}\,.
\label{Omechi2}
\ee
Since $\Omega_{\chi 0} \le \Omega_{M0}$ and 
$(a_*+a_{\rm eq})/a_*>1$, the inequality (\ref{Omechi2}) 
holds for $\beta_{m*} \ll 1$.

Expressing the coefficient in front of $\delta_{\chi {\rm N}}$ in Eq.~(\ref{delm3})
as 
\be
\omega_\chi^2=c_s^2 \frac{k^2}{a^2}-4\pi G \rho_{\chi}
=\frac{2m^2 k^2+\lambda a^2 \rho_{\chi}}{2m^2 (4m^2 a^2+k^2)} 
\frac{k^2}{a^2}-4\pi G \rho_{\chi}\,,
\ee
the behavior of $\omega_\chi^2$ is different depending on $k$. 
During the time interval $a_*<a<a_{\rm E}$, its scale dependence 
is given by 
\ba
\omega_\chi^2 \simeq 
\begin{cases}   
   \dfrac{k^4}{a^2(4m^2 a^2+k^2)}\,, & \qquad (k>k_I)\,, \\
   \dfrac{\lambda \rho_{\chi} k^2}{2m^2(4m^2 a^2+k^2)}\,, 
   & \qquad (k_S<k<k_I)\,, \\
   -4\pi G \rho_{\chi}\,, 
   & \qquad (k<k_S)\,.
   \label{omese}
\end{cases}
\ea
For $k>k_I$ the quantum pressure suppresses the growth of $\delta_{\chi{\rm N}}$, 
while, for $k<k_S$, $\delta_{\chi{\rm N}}$ grows in the standard manner 
by the gravitational instability.
In the intermediate wavenumber $k_S<k<k_I$, the self-coupling 
contribution to $\omega_\chi^2$ dominates 
over both quantum pressure 
and gravitational instability.
Taking into account the contribution of quantum pressure 
to $c_s^2$ in the regime $k_S<k<k_I$ with $k/a \ll m$, 
we find that $\omega_\chi^2$ takes a minimum value 
\be
\omega_{\chi,{\rm min}}^2=
-4\pi G \rho_{\chi}
-\frac{\lambda^2 \rho_{\chi}^2}{64m^6}
=-4\pi G \rho_{\chi} \left( 
1-\frac{1}{4} \beta_U \mu \right) \,,
\label{omechi}
\ee
at 
\be
k_{\rm min}=\frac{a\sqrt{|\lambda| \rho_{\chi}}}{2m}
=\frac{k_I}{\sqrt{2}}\,.
\label{kmin}
\ee
Even if $|\beta_U| \ll 1$, the term $-\beta_U \mu/4$ can be 
of order 1, so that it is compatible with the
gravitational instability term in $\omega_{\chi,{\rm min}}^2$.
In the presence of other perturbations like $\delta_{c{\rm N}}$, 
the last term on the left hand-side of Eq.~(\ref{delm3}) 
also works as a source term for the gravitational instability 
of $\delta_{\chi{\rm N}}$.

For $a>a_{\rm E}$, the quantity $\omega_{\chi}^2$ has 
the following scale dependence
\ba
\omega_\chi^2 \simeq 
\begin{cases}   
\dfrac{k^4}{a^2(4m^2 a^2+k^2)}\,, & \qquad (k>k_J)\,, \\
-4\pi G \rho_{\chi}\,, & \qquad (k<k_J)\,.
\end{cases}
\ea
In this regime, there is no range of $k$ in which the self-coupling 
potential affects the growth of $\delta_{\chi{\rm N}}$.

\subsection{Evolution of density contrasts for ultra-light BEC}

For the mass range $m<7 \times 10^{-28}$~{\rm eV}, 
the scalar field starts to oscillate after matter-radiation equality. 
In this case, the scales of Laplacian instabilities discussed above 
can be as large as those relevant to the observations of CMB and 
matter power spectra in the linear regime.
In the following, we study the evolution of perturbations 
in this ultra-light mass range.
On using the approximation $a_* \gg a_{\rm eq}$ for 
$\beta_{m*}$ in Eq.~(\ref{Omechi}), we have
\be
a_* \simeq \left( \frac{H_0^2 \Omega_{M0}}{m^2 \beta_{m*}} 
\right)^{1/3}\,.
\ee
The inequality (\ref{Omein}) yields
\be
\sqrt{\frac{2}{3} \frac{1}{ \beta_{m*}} \frac{\Omega_{M0}}
{\Omega_{\chi 0}}}
< |\mu| \ll \frac{2}{3} \frac{1}{ \beta_{m*}} \frac{\Omega_{M0}}
{\Omega_{\chi 0}}\,.
\label{muup}
\ee
Under this upper bound of $|\mu|$, the quantity 
$-\beta_U \mu/4$ in Eq.~(\ref{omechi}) is in the range
\be
-\frac{1}{4} \beta_U \mu  \ll \frac{\beta_U}{6 \beta_{m*}}
\frac{\Omega_{M0}}{\Omega_{\chi 0}}\,.
\label{bU}
\ee
If the $\chi$ field is responsible for all DM 
($\Omega_{\chi 0}=\Omega_{M0}$), 
the right hand-side of Eq.~(\ref{bU}) is given by 
$\beta_U/(6\beta_{m*})$. 
At $a=a_*$, it is natural to consider the situation in which   
$\beta_{U*}$ is of the similar order to $\beta_{m*}$ and 
hence the quantity $\beta_{U*}/(6\beta_{m*})$ is at most 
of order 1. For $a>a_*$, $\beta_U$ decreases in proportion 
to $a^{-3}$, so the term $-\beta_U \mu/4$ is constrained 
to be much smaller than 1. 
This means that, for $\Omega_{\chi 0}=\Omega_{M0}$, 
the self-coupling term in $\omega_{\chi,{\rm min}}^2$ 
does not significantly dominate over the gravitational 
instability term $-4\pi G \rho_{\chi}$.

If $\Omega_{\chi 0} \ll \Omega_{M0}$, then the term 
$-\beta_U \mu/4$ can be larger than 1 due to the large upper limit on the 
right hand-side of Eq.~(\ref{bU}).
For $\mu=0$, the likelihood analysis using the CMB 
and galaxy clustering data in the ultra-light axion mass range 
$10^{-32}~{\rm eV} \le m \le 10^{-25.5}~{\rm eV}$ 
showed that the $\chi$-field density parameter is constrained to be 
$\Omega_{\chi 0}/\Omega_{M0} \le 0.05$  \cite{Hlozek:2014lca}. 
This small value of $\Omega_{\chi 0}$ relative to the total 
DM density is attributed to the fact that the quantum 
pressure suppresses the growth of the BEC density contrast 
for $k>k_J$. Even if the self-interaction were to enhance 
$\delta_{\chi {\rm N}}$ at some particular scales, 
the quantum pressure still leads to the suppression of 
$\delta_{\chi {\rm N}}$ on scales relevant to the observational 
range of CMB and linear matter power spectrum. 
Hence the fact that $\Omega_{\chi 0}$ needs to be much smaller 
than $\Omega_{M0}$ for the above ultra-light mass region 
should not be modified.

Let us consider the mass range $m<7 \times 10^{-28}$~{\rm eV} 
together with the condition $\Omega_{\chi 0} \ll \Omega_{M0}$ 
in the following discussion. 
In this case, we need to take CDM and baryons 
into account to study the evolution of perturbations 
after matter dominance.
For $a>a_* \gg a_{\rm eq}$, it is unnecessary to distinguish between 
CDM and baryons, in that both perturbations satisfy Eqs.~(\ref{pereq3d}) 
and (\ref{pereq4d}) with Eq.~(\ref{cIwI}). 
We define the CDM-baryon density contrast
$\delta_{m{\rm N}}=(\Omega_c \delta_{c{\rm N}}+\Omega_b 
\delta_{b{\rm N}})/(\Omega_c+\Omega_b)$, with today's 
density parameter $\Omega_{m0} \equiv \Omega_{c0}+\Omega_{b0}=
\Omega_{M 0}-\Omega_{\chi 0}$. 
In Eq.~(\ref{delm3}), there exists the contribution of order 
$-4\pi G \rho_m \delta_{m{\rm N}}$ (where $\rho_m=\rho_c+\rho_b$), 
besides the term proportional to $\delta_{\chi{\rm N}}$. 
We recall that the coefficient of $\delta_{\chi{\rm N}}$ takes  
the minimum value (\ref{omechi}) at the wavenumber (\ref{kmin}).
The ratio between 
$\omega_{\chi,{\rm min}}^2 \delta_{\chi {\rm N}}$ 
and $-4 \pi G \rho_{m} \delta_{m {\rm N}}$ is given by 
\be
\frac{\omega_{\chi,{\rm min}}^2 \delta_{\chi {\rm N}}}
{-4 \pi G \rho_{m} \delta_{m {\rm N}}}
=\left( 1-\frac{1}{4} \beta_U \mu \right) 
\frac{\Omega_{\chi 0}}{\Omega_{m0}} 
\frac{\delta_{\chi {\rm N}}}{\delta_{m {\rm N}}}\,.
\label{rchi}
\ee
Applying the upper limit of $|\mu|$ in Eq.~(\ref{muup}) 
to the second term on the right hand-side of Eq.~(\ref{rchi}), it follows that 
\be
r_U \equiv
-\frac{1}{4}\beta_U \mu \frac{\Omega_{\chi 0}}{\Omega_{m0}}
\frac{\delta_{\chi {\rm N}}}{\delta_{m {\rm N}}} \ll
\frac{\beta_U}{6 \beta_{m*}}
\frac{\Omega_{M0}}{\Omega_{m 0}}
\frac{\delta_{\chi {\rm N}}}{\delta_{m {\rm N}}}
\equiv r_{U{\rm max}}\,.
\label{bucon}
\ee
Since $\Omega_{m 0} \simeq \Omega_{M 0}$ for $\Omega_{\chi 0} \ll \Omega_{M0}$,  
the upper limit of Eq.~(\ref{bucon}) reduces to 
$r_{U{\rm max}} \simeq \beta_U/(6\beta_{m*})\delta_{\chi {\rm N}}/\delta_{m {\rm N}}$. 
Provided that $\beta_{U*}$ is the similar order to $\beta_{m*}$ 
and that $\delta_{\chi {\rm N}}$ is initially of the same order as 
$\delta_{m{\rm N}}$ (which is the case for adiabatic initial 
conditions discussed below),
the term $r_{U}$ is suppressed to be smaller than 1. 
Moreover, the first term $(\Omega_{\chi 0}/\Omega_{m0})
(\delta_{\chi {\rm N}}/\delta_{m {\rm N}})$ on the right hand-side 
of Eq.~(\ref{rchi}) is much less than 1.
Then the evolution of $\delta_{\chi {\rm N}}$ is dominated 
by the gravitational instability term $-4 \pi G \rho_{m} \delta_{m {\rm N}}$ 
rather than the negative Laplacian term 
$c_s^2 (k^2/a^2) \delta_{\chi {\rm N}}$.
This suggests that the self-coupling would not lead to the strong 
Laplacian instability of $\delta_{\chi {\rm N}}$.
For perturbations deep inside the Hubble radius, the CDM-baryon density contrast 
$\delta_{m{\rm N}}$ approximately obeys Eq.~(\ref{delNeq}). 
Since the term $-4 \pi G \rho_{m} \delta_{m {\rm N}}$ also dominates over 
$-4\pi G \rho_{\chi}(1+c_s^2) \delta_{\chi {\rm N}}$, the evolution 
of $\delta_{m{\rm N}}$ should be hardly modified by $\delta_{\chi {\rm N}}$.

To confirm these properties, we numerically integrate the perturbation equations 
of motion for the mass $m=10^{-29}$~eV without using the sub-horizon 
approximation explained in Sec.~\ref{anasec}. 
Besides CDM and baryons, we take DE into account 
as the cosmological constant and take today's density parameters 
$\Omega_{\chi 0}=0.01$, $\Omega_{m 0}=0.30$, and 
$\Omega_{{\rm DE}0}=0.69$. For the onset of BEC formation, 
we choose the moment $m=10H_*$, i.e., $\beta_{m*}=0.01$, 
so that $a_*=8.74 \times 10^{-3}$. 
Then the self-coupling region (\ref{Omein}) is given by 
$46.2<|\mu| \ll 2.14 \times 10^3$. 
We choose $\mu=-250$ in our numerical simulation, 
in which case $\beta_{U*}=0.12$. 
This order of $\beta_{U*}$, which is 10 times as large as $\beta_{m*}$,  
should be regarded as a maximum
for the validity of nonrelativistic BEC description. 
The minimum and maximum wavenumbers 
in the yellow shaded region of Fig.~\ref{fig2} at $a=a_*$ 
are given, respectively, by $k_{S*}=1.22 \times 10^{-3}$~Mpc$^{-1}$ 
and $k_{I*}=6.62 \times 10^{-3}$~Mpc$^{-1}$.
For the density contrasts and velocity potentials, 
we choose the adiabatic initial conditions
\be
\delta_{\chi {\rm N}}(a_*)=\delta_{c {\rm N}}(a_*)=
\delta_{b {\rm N}}(a_*)\,,\qquad 
v_{\chi {\rm N}}(a_*)=v_{c {\rm N}}(a_*)=
v_{b {\rm N}}(a_*)\,,
\label{adi}
\ee
together with $\Phi(a_*)=-2.8 \times 10^{-5}$ and 
$\dot{\Phi}(a_*)=0$.

\begin{figure}[h]
\begin{center}
\includegraphics[height=3.15in,width=3.5in]{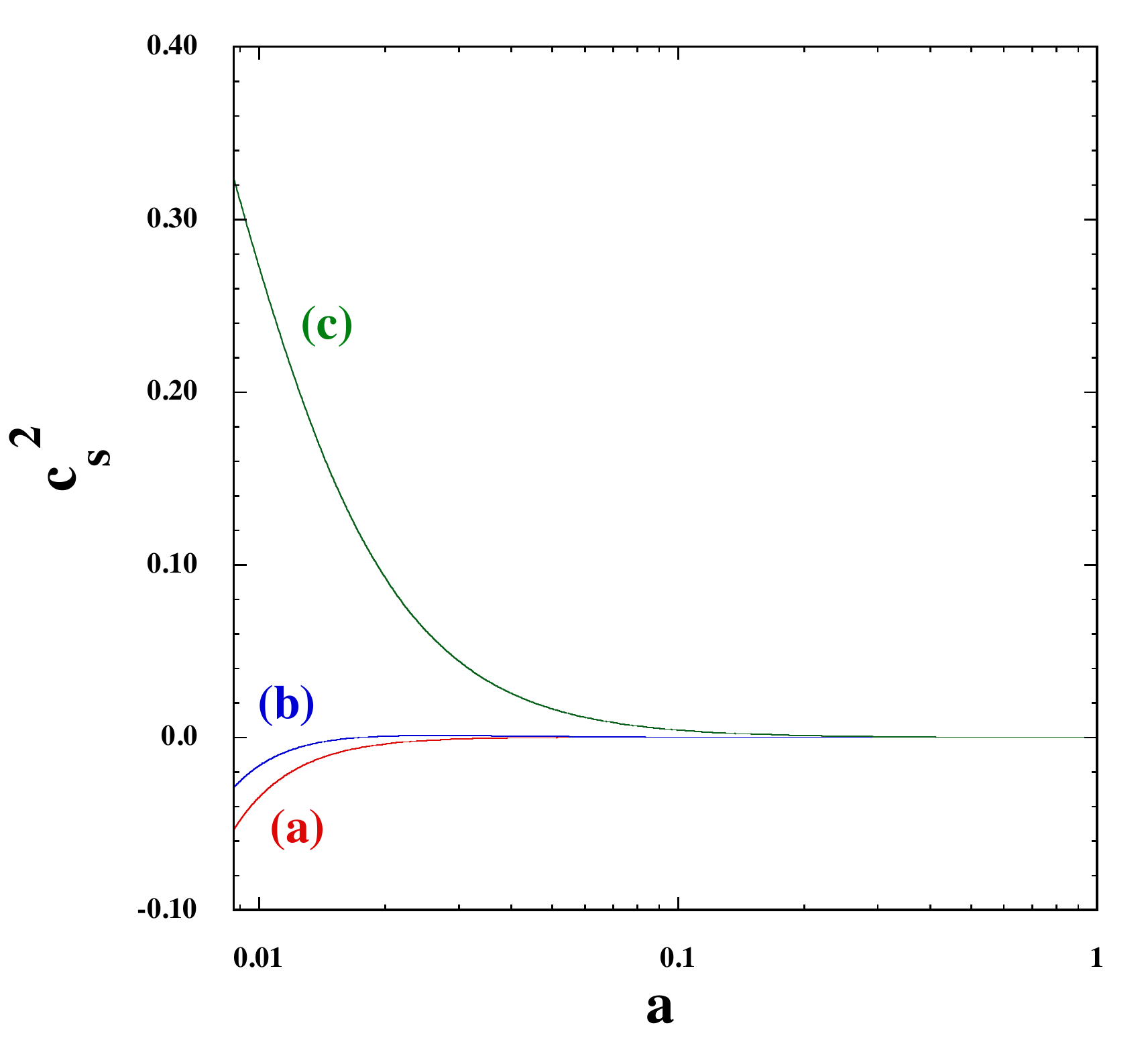}
\includegraphics[height=3.1in,width=3.4in]{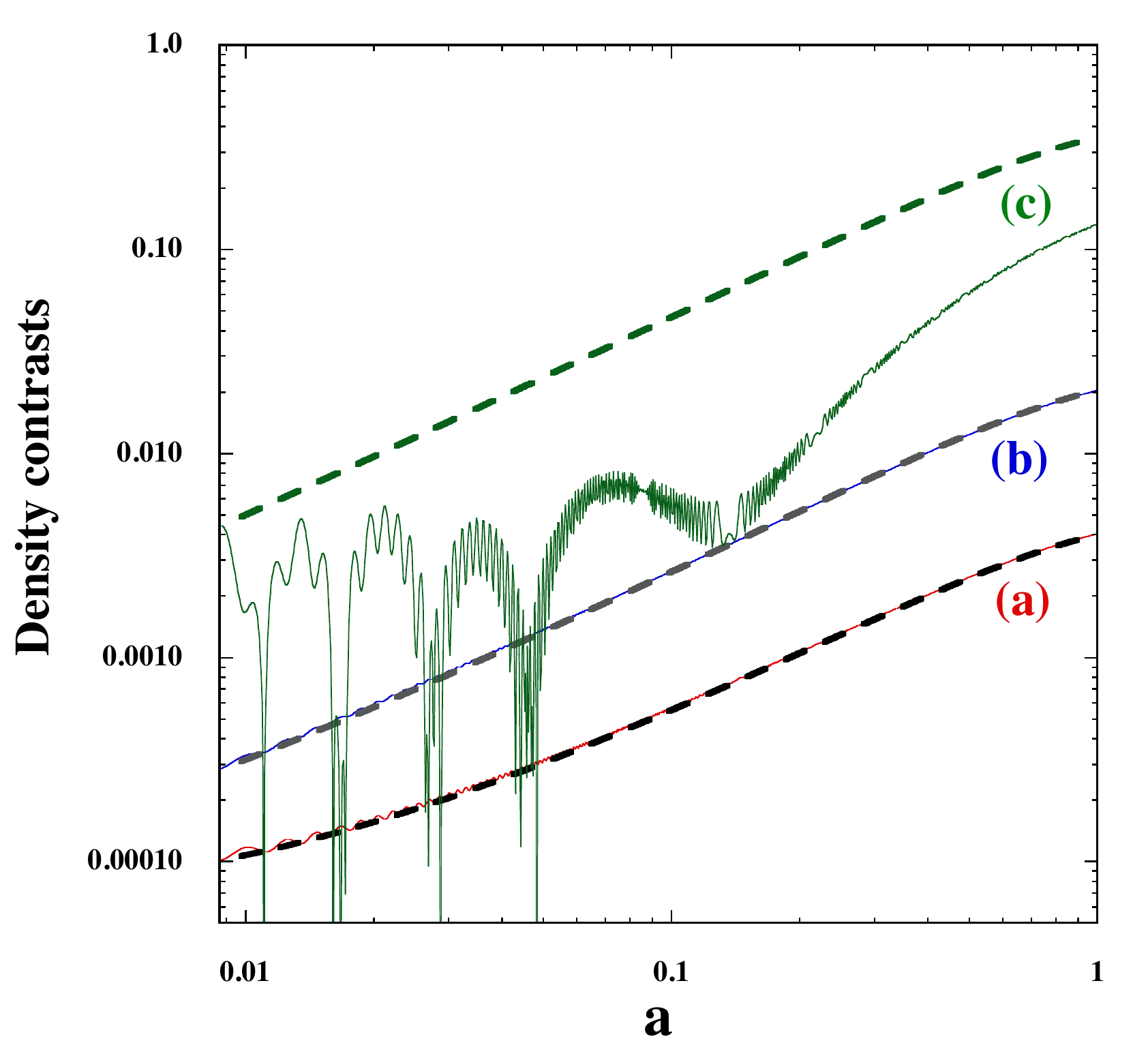}
\end{center}
\caption{\label{fig3}
(Left) 
Evolution of the BEC sound speed squared $c_s^2$ versus 
the scale factor $a$ for $m=10^{-29}$~eV and $\mu=-250$. 
The red, blue, and green lines correspond to the modes
(a) $k=2.05 \times 10^{-3}$~Mpc$^{-1}$, 
(b) $k=4.68 \times 10^{-3}$~Mpc$^{-1}$, and 
(c) $k=2.05 \times 10^{-2}$~Mpc$^{-1}$, respectively. 
The initial conditions of density parameters at $a_*=8.74 \times 10^{-3}$
are chosen to realize today's values $\Omega_{\chi 0}=0.01$, 
$\Omega_b=0.05$, $\Omega_c=0.25$, 
and $\Omega_{\rm DE}=0.69$.
(Right) 
The thin red, blue, and green lines show the evolution of $\delta_{\chi {\rm N}}$ 
for the same wavenumbers (a), (b), (c) as those used in the left panel.
The thick dashed black, grey, and green lines correspond to 
the growth of CDM-baryon density contrast 
$\delta_{m{\rm N}}$ for the wavenumbers (a), (b), (c), respectively.} 
\end{figure}

In the left panel of Fig.~\ref{fig3} we plot the evolution of 
$c_s^2$ for three different values of $k$. 
In cases (a) and (b) the wavenumbers are 
$k=2.05 \times 10^{-3}$~Mpc$^{-1}$ and $k=4.68 \times 10^{-3}$~Mpc$^{-1}$, 
respectively, which are both in the region $k_{S*}<k<k_{I*}$.
In these cases, $c_s^2$ is initially negative up to a critical 
scale factor $a_{\rm I}$, after which the sign of $c_s^2$ 
changes to be positive. 
For a given mode $k$, this critical scale factor 
is determined by the condition $k_I=k$, such that 
\be
a_{\rm I}=3.34 \times 10^{-7} h^2 \Omega_{\chi 0} |\mu|
\left( \frac{1~{\rm Mpc}^{-1}}{k} \right)^2\,.
\ee
In cases (a) and (b) we have $a_{\rm I}=9.10 \times 10^{-2}$ and 
$a_{\rm I}=1.74 \times 10^{-2}$, respectively, which are in agreement 
with the numerical results of Fig.~\ref{fig3}. 
We note that the wavenumber (\ref{kE}) 
at $k_J=k_S=k_I$ is $k_{\rm E}=3.77 \times 10^{-3}$~Mpc$^{-1}$ 
with $a_{\rm E}=2.69 \times 10^{-2}$.
In case (a), which corresponds to $k<k_{\rm E}$, 
$a_{\rm I}$ is larger than $a_{\rm E}$. 
However, the parameter space in which $c_s^2 k^2/a^2$ dominates 
over $-4\pi G \rho_{\chi}$ is limited to the yellow shaded 
region in Fig.~\ref{fig2}. 
For $k<k_{\rm E}$, this dominance occurs for $a_*<a<a_{\rm S}$, where 
$a_{\rm S}~(<a_{\rm E})$ is determined by the condition $k_S=k$, i.e., 
\be
a_{\rm S}=2.12 \times 10^{3} h^{-1} \sqrt{|\mu|} 
\frac{H_0}{m} \frac{k}{1~{\rm Mpc}^{-1}}\,.
\ee
For $k_{\rm E}<k<k_{I*}$, 
the region of self-coupling dominance is
in the interval $a_*<a<a_{\rm I}$, where $a_{\rm I}$ 
is smaller than $a_{\rm E}$.  
The wavenumber in case (b) belongs to this range. 
In case (c) of Fig.~\ref{fig3} the wavenumber is 
larger than $k_{I*}$, so $c_s^2$ is always positive for $a>a_*$.

In the right panel of Fig.~\ref{fig3}, we plot the evolution of 
$\delta_{\chi {\rm N}}$ as well as $\delta_{m {\rm N}}$ 
for the three wavenumbers same as those used in the left panel.
As we see in Eq.~(\ref{pereq5d}), for increasing $k$, the initial amplitudes 
of density contrasts tend to be larger.
In particular the wavenumbers in cases (a) and (b) are initially close to 
the Hubble radius (${\cal K}(a_*)=1.5$ and ${\cal K}(a_*)=3.4$ respectively), 
so we numerically solve the full perturbation equations of motion 
without resorting to the sub-horizon approximation explained 
in Sec.~\ref{anasec}. Indeed, using the approximate 
Eqs.~(\ref{delm3}) and (\ref{delNeq}) for these modes 
gives rise to some difference in comparison to 
the full numerical results.

In Fig.~\ref{fig3}, we observe that the evolutions of $\delta_{\chi {\rm N}}$ 
in cases (a) and (b) are practically identical to that of $\delta_{m {\rm N}}$. 
This means that the negative value of $c_s^2$ induced by 
the self-coupling does not lead to the additional enhancement 
of $\delta_{\chi {\rm N}}$ besides the gravitational instability. 
In case (b), the wavenumber corresponds to $k=k_{{\rm min}*}=k_{I*}/\sqrt{2}$, 
so $\omega_\chi^2$ takes the minimum value (\ref{omechi}).
At $a=a_*$, the quantities in Eq.~(\ref{bucon}) are given by $r_{U*}=0.25$ 
and $r_{U{\rm max}*}=2.1$. 
Since $r_{U*}$ is less than the order 1, the self-coupling effect
on the growth of $\delta_{\chi {\rm N}}$ is suppressed relative to 
the gravitational instability term $-4\pi G \rho_m \delta_{m{\rm N}}$ 
arising from CDM and baryons. 
Due to the decrease of $r_{U}$ in time, the contribution of $r_U$ to the 
ratio (\ref{rchi}) tends to be weaker at late times.
Not only in cases (a) and (b), but also for the wavenumbers 
in the range $k_{S*}<k<k_{I*}$, 
we find that $\delta_{\chi {\rm N}}$ is hardly subject 
to the Laplacian instability.
This property is mostly attributed to the upper limit of $|\mu|$ 
given in Eq.~(\ref{muup}).
We recall that the baryon-CDM density contrast 
$\delta_{m{\rm N}}$ acquires the BEC sound speed squared 
$c_s^2$ in Eq.~(\ref{delNeq}). Since $|c_{s}^2|$ decreases in time 
with the initial value smaller than the order 1, the effect of $c_s^2$ 
on the evolution of $\delta_{m{\rm N}}$ 
can be negligible especially for $\Omega_{\chi 0} \ll \Omega_{m0}$.

In case (c) of Fig.~\ref{fig3}, the wavenumber is in the range $k>k_{I*}$ 
and hence the quantum pressure suppresses the growth 
of $\delta_{\chi {\rm N}}$ at early times.
In the late epoch, however, the term $-4\pi G \rho_m \delta_{m{\rm N}}$
dominates over the positive Laplacian term 
$c_s^2 (k^2/a^2) \delta_{\chi{\rm N}}$ in Eq.~(\ref{delm3}). 
Then, $\delta_{\chi{\rm N}}$ starts to grow at some point to 
catch up with $\delta_{m{\rm N}}$. 
For increasing $k$, the initial epoch during which 
$\delta_{\chi {\rm N}}$ does not grow by the quantum pressure 
tends to be longer, so today's value of $\delta_{\chi {\rm N}}$ 
is more significantly suppressed in comparison to 
$\delta_{m {\rm N}}$. 
Thus, for the modes $k>k_{I*}$, the self-coupling does not affect 
the dynamics of perturbations, but the quantum pressure plays 
an important role to modify the gravitational clustering of
$\delta_{\chi {\rm N}}$.

Finally, we comment on the analysis of 
Refs.~\cite{Zhang:2017flu,Cedeno:2017sou,Zhang:2017dpp,Arvanitaki:2019rax}, 
in which the authors studied the evolution of an axion density contrast 
for the potential (\ref{Vpo}). They considered the axion mass in the range 
$m=10^{-22}$\,eV\,$\sim\,$10$^{-21}$~eV and showed that the self-coupling 
with a large field misalignment can induce instabilities of the axion 
density contrast for particular wavenumbers around $k=10$~Mpc$^{-1}$. 
In this case, the axion starts to oscillate long before matter-radiation 
equality, so it is necessary to take the radiation perturbation into account. 
For the large misalignment, the axion density contrast can be
enhanced by parametric resonance between 
the onset of field oscillation 
and the formation of BEC ($a_{\rm osc}< a<a_*$).
Since our nonrelativistic BEC description amounts to averaging 
over oscillations in the regime $a>a_*$, it does not accommodate 
the phenomenon of parametric resonance during such a transient epoch. 
What we showed in this paper is that, after the BEC formation, 
the Laplacian instability associated with negative values of 
$c_s^2$ induced by the attractive self-interaction is no longer 
effective for the mass range $m<7 \times 10^{-28}$~{\rm eV}.
It will be of interest to explore whether the resonant instability is also 
present for such a ultra-light mass range 
to probe signatures of axion perturbations on scales 
larger than 10~Mpc.

\section{Conclusions}
\label{concludesec}

In this paper, we provided a general framework for studying the evolution of 
cosmological perturbations for an ultra-light complex scalar field $\chi$ 
in a state of the nonrelativistic BEC.
Using the Madelung representation (\ref{chiMa}),
we expressed the continuity and Euler equations as well as the 
gravitational field equations in general relativistic, covariant forms. 
We also included other matter sources like CDM as perfect fluids 
and explicitly showed their difference from the BEC in a covariant 
manner. On the FLRW background, the regime in which 
the nonrelativistic BEC is formed is characterized by the 
conditions (\ref{mcon}).

In Sec.~\ref{persec}, we derived the full linear perturbation equations 
of motion for the line element (\ref{permet}) containing four scalar 
perturbed variables $\alpha$, $B$, $\zeta$, $E$.
The BEC matter perturbation $\delta \rho_{\chi}$ and its velocity 
potential $v_{\chi}$ obey Eqs.~(\ref{pereq1}) and (\ref{pereq2}), 
respectively. In comparison to the continuity and Euler 
Eqs.~(\ref{pereq3}) and (\ref{pereq4}) of nonrelativistic perfect 
fluids with $w_I=c_I^2=0$, there exists the perturbation $\delta \beta$ 
given by Eq.~(\ref{dbeta}) which contains the effects of quantum 
pressure and BEC self-interactions. 
Metric perturbations are coupled to the energy-momentum 
tensors of both BEC and perfect fluids through the Einstein Eq.~(\ref{Einf}). 
Our perturbation equations can be applied to any choice of gauges 
depending on the problem at hand. 
We showed that all the perturbed equations can be expressed 
in terms of the gauge-invariant variables introduced 
in Eq.~(\ref{delphiN}).

In Sec.~\ref{subsec}, we used the quasi-static approximation for perturbations 
deep inside the Hubble radius. Even though the background value of $\beta$ 
vanishes, its gauge-invariant perturbation $\delta \beta_{\rm N}$ contains an oscillating mode 
$\delta \beta_{\rm N}^{(\rH)}$ with the frequency associated with 
the field mass $m$. There is also the special solution 
$\delta \beta_{\rm N}^{(\rs)}$ sourced by the density contrasts 
$\delta_{\chi {\rm N}}$ and $\delta_{I {\rm N}}$, see Eq.~(\ref{delbs}). 
Taking the average of Eq.~(\ref{delmos}) over the Hubble time scale 
amounts to ignoring $\delta \beta_{\rm N}^{(\rH)}$ relative to $\delta \beta_{\rm N}^{(\rs)}$. 
Then, the BEC density contrast $\delta_{\chi {\rm N}}$ obeys the approximate equation of the form 
(\ref{delm3}), with the effective sound speed squared (\ref{cs}). 
This value of $c_s^2$, which contains the contributions of quantum pressure 
and BEC self-interactions, reproduces those known in the literature 
in the regime $k/a \ll m$ for the two-body self-interacting potential
$U(\rho)=\lambda \rho^2/4$. 
We also showed that the density contrasts $\delta_{I{\rm N}}$ in the 
perfect-fluid sector are affected by the BEC sound speed 
in the form (\ref{delNeq}) through the gravitational interaction.
We numerically solved the perturbation equations of $\delta_{\chi{\rm N}}$
during the matter era dominated by the BEC energy density without self-interactions 
and found that the approximate Eq.~(\ref{delm3}) can be trustable except for 
the wavenumbers initially around the Hubble radius.

In Sec.~\ref{selfsec}, we investigated the effect of the BEC two-body self-interacting 
potential $U(\rho)=\lambda \rho^2/4$ on the dynamics of linear perturbations. 
We also included CDM, baryons, and the cosmological constant to 
discuss the dynamics after the onset of matter dominance. 
We studied the evolutions of $\delta_{\chi {\rm N}}$ and the CDM-baryon density 
contrast $\delta_{m{\rm N}}$ in the ultra-light 
mass range $m<7 \times 10^{-28}$~eV, which was unexplored before
in the presence of self-interactions. The BEC sound speed squared 
$c_s^2$ can be negative by its self-interactions for some particular 
range of scales relevant to the CMB and large-scale structure measurements.
In this ultra-light mass range, the BEC cannot be all DM due to the suppression 
of the matter power spectrum for $k>k_J$. 
Under the constraint $\Omega_{\chi 0} \ll \Omega_{M0}$, the negative Laplacian 
term $c_s^2 k^2/a^2$ induced by the self-coupling can dominate over
the other term $-4\pi G \rho_{\chi}$ appearing as the 
coefficient of $\delta_{\chi {\rm N}}$ in Eq.~(\ref{delm3}). 
However, the gravitational instability term $-4\pi G \rho_m \delta_m$ 
arising from the CDM-baryon density contrast $\delta_m$ in Eq.~(\ref{delm3}) 
overwhelms $c_s^2 k^2/a^2$ for the adiabatic initial conditions (\ref{adi}).
Numerically, we confirmed that, even when $c_s^2$ is negative, 
the self-coupling hardly induces 
the Laplacian instabilities of $\delta_{\chi {\rm N}}$ and $\delta_{m{\rm N}}$
besides their gravitational instabilities. 

We have thus shown that the BEC self-coupling is almost ineffective 
to modify the dynamics of linear density perturbations at least 
in the regime where the nonrelativistic BEC description is valid. 
In the context of axions, there is a transient epoch between 
the moments at which the axion 
starts to oscillate ($m \simeq 3H$) and when the BEC is formed ($m \gg H$).
To deal with such a transition including the phenomenon of 
parametric resonance, we need to consider perturbations of 
the bosonic field itself and relate them with its density contrast and 
velocity potential. After the BEC formation, those perturbations should be 
matched with the solutions derived in this paper. 
The detailed study about the evolution of inhomogeneities including 
such a transient epoch with the various axion mass range
deserves for a future separate work.

\section*{Acknowledgements}

ST is supported by the Grant-in-Aid for Scientific Research Fund 
of the JSPS No.\,19K03854.


\end{document}